\documentclass[preprint,showpacs,showkeys,floats,floatfix,nofootinbib]{revtex4}

\usepackage{amsbsy}
\usepackage{amssymb}
\usepackage{amsmath}
\input{epsf}
\usepackage{graphicx}
\usepackage{amssymb}
\usepackage{bm}
\usepackage{xcolor,cancel}

\def\spose#1{\hbox to 0pt{#1\hss}}

\def\lta{\mathrel{\spose{\lower 3pt\hbox{$\mathchar"218$}}
     \raise 2.0pt\hbox{$\mathchar"13C$}}}
\def\gta{\mathrel{\spose{\lower 3pt\hbox{$\mathchar"218$}}
     \raise 2.0pt\hbox{$\mathchar"13E$}}}

\def\beq{\begin{equation}}
\def\eeq{\end{equation}}
\def\bea{\begin{eqnarray}}
\def\eea{\end{eqnarray}}

\def\x{{\rm x}}
\def\y{{\rm y}}
\def\z{{\rm z}}
\def\n{{\rm n}}

\def\s{{\rm s}}

    \def\X{{\rm x}}
    \def\Y{{\rm y}}
\newcommand{\A}{{\cal A}}
\newcommand{\B}{{\cal B}}

\def\fd{\xi}
\def\lefx{{\cal L}_{\fd_\x}}
\def\lefy{{\cal L}_{\fd_\y}}

\def\RXYa{ R^{\x \y}_a}
\def\RYXa{ R^{\y \x}_a}

\def\DXab{{S}^\x_{a b}}
\def\DXba{{S}^\x_{b a}}

\def\DXuab{{S}_\x^{a b}}

\def\RHXYa{{ r}^{\x \y}_a}
\def\RHYXa{{ r}^{\y \x}_a}

\def\DXYab{{s}^{\x \y}_{a b}}
\def\DXYba{{s}^{\x \y}_{b a}}

\def\DYXba{{s}^{\y \x}_{b a}}

\def\DXYuab{{s}_{\x \y}^{a b}}

\def\rHXYa{{\cal R}^{\x \y}_a}
\def\rHYXa{{\cal R}^{\y \x}_a}

\def\dXYab{{\cal S}^{\x \y}_{a b}}

\def\dYXba{{\cal S}^{\y \x}_{b a}}

\def\dXYusab{{\cal S}_{\x \y}^{a b}}

\def\hxab{g^{A B}_\x}
\def\hxba{g^{B A}_\x}
\def\hxde{g^{D E}_\x}

\def\hyab{g^{A B}_{\y}}

\def\hyde{g^{D E}_{ \y}}

\def\hxyab{g^{A B}_{\x \y}}

\def\hyxba{g^{B A}_{\y \x}}
\def\hxyde{g^{D E}_{\x \y}}
\def\hxylbabrb{g^{[A B]}_{\x \y}}

\def\rtotxa{R_a^\x}

\def\dtotxba{D^\x_{b a}}
\def\dtotxab{D^\x_{a b}}

\def\dtotxabu{D_\x^{a b}}

\begin{document}
\title{A covariant action principle for dissipative fluid dynamics: From formalism to 
fundamental physics}

\author{N.~Andersson}
\email{na@maths.soton.ac.uk}
\affiliation{Mathematical Sciences \& STAG Research Centre, University of 
Southampton, Southampton SO17 1BJ, UK}
 
\author{G.~L.~Comer}
\email{comergl@slu.edu}
\affiliation{Department of Physics, Saint Louis University, St.~Louis, MO, 63156-0907, 
USA}

\date{\today}

\begin{abstract}
We present a new variational framework for dissipative general relativistic fluid 
dynamics. The model extends the convective variational principle for multi-fluid systems 
to account for a range of dissipation channels. The key ingredients in the construction 
are i) the use of a lower dimensional matter space for each fluid component, and ii) an 
extended functional dependence for the associated volume forms. In an effort to make 
the concepts clear, the formalism is developed step-by-step with  model examples 
considered at each level. Thus we consider a model for heat flow, derive the relativistic 
Navier-Stokes equations and discuss why the individual dissipative stress tensors need 
not be spacetime symmetric. We argue that the new formalism, which notably does not 
involve an expansion away from an assumed equilibrium state, provides a conceptual 
breakthrough in this area of research. We also provide an ambitious list of directions in 
which one may want to extend it in the future.  This involves an exciting set of problems, 
relating to both applications and foundational issues.
\end{abstract}


\maketitle
\section{Introduction}

The marriage between the  theory of general relativity and thermodynamics is known to 
be conceptually challenging. On the one hand, the breakthrough associated with 
Einstein's theory was due to an understanding of the covariance of physical laws, 
leading to the concept of spacetime and an emphasis on the observer's role in making 
measurements. On the other hand, thermodynamics identifies a specific direction of time 
associated with the second law and the inevitable increase of entropy. Hence, it is not 
surprising that the development of models for non-equilibrium thermodynamical systems 
consistent with the tenets of general relativity remains a topical problem 
\cite{israel81:_rel_thermo,1989LNM..1385..152I,2013PhRvD..87h4055R}. A workable 
model for dissipative fluid dynamics is required for a range of applications, from 
astrophysics and cosmology (perhaps particularly in the context of numerical 
simulations \cite{lrr-2003-7,lrr-2008-7}) to the description of hot dense plasmas for 
colliders like RHIC and the LHC \cite{2004PhRvC..69c4903M,2013arXiv1301.5893G,
2013arXiv1301.2826H}. We also need to make progress on foundational issues. In 
particular, one should clarify the link between phenomenological macro-models and the 
relevant processes on the micro-scale 
\cite{stewart77:_rel_kin_theor,Israel79:_kintheo2,2011PhRvE..84b6315F,2011CQGra..28q5003C}, and understand the intimate relation between the flow of time and a 
system's evolution towards dynamical and thermodynamical equilibrium.  

\subsection{Motivation and scope}

The aim of  this paper is to demonstrate, mainly as a  proof-of-principle, that the 
dynamics of a dissipative multifluid system can be obtained from a constrained 
variational principle. This is an  exciting result which promises to lead to significant 
progress in this problem area.

The premise of our discussion may seem at odds with the conventional wisdom, 
according to which action principles --- expressed as an integral of a 
Lagrangian, whose local extrema satisfy the equations of motion, subject to 
well-posed boundary constraints --- do not exist for dissipative systems. However, given 
the foundational nature of the problem it is quite natural to consider it and, in fact, a 
number of more or less successful attempts to make progress can be found in the 
literature. A common approach has been to combine a variational model for the non-
dissipative problem with an argument that constrains  the entropy production, often 
involving Lagrange multipliers (see \cite{1994PhR...243..125I} for a review and 
\cite{1975AcMec..23...17D,1980JPhA...13..431D,1982RSPSA.381..457M,1986AmJPh..54..997K,1986IJNLM..21..489V,1991PhLA..155..223H,Chien,2007EPJD...44..407N,2012PThPh.127..921F} 
for samples of the literature). The variational model we will develop is conceptually 
different. The conservative constraints on the system are built into the variation itself and 
the model does not involve (at least not in the first instance) an expansion away from 
equilibrium (in contrast to the  celebrated ``second order'' model of Israel and Stewart 
from the 1970s  \cite{Israel76:_thermo,Israel79:_kintheo1,Israel79:_kintheo2}  or, 
indeed, any model based on a  derivative expansion from the beginning). This means 
that the new description remains valid (at least formally) also for systems far away from 
equilibrium, and hence it provides a promising framework for the exploration of 
nonlinear thermodynamical evolution and associated irreversible phenomena.  This is a 
problem area where a number of challenging issues remain to be resolved, involving for 
example maximum versus minimum entropy production for non-equilibrium systems 
\cite{1980ARPC...31..579J,2003JPhA...36..631D,2006PhR...426....1M,2010PhRvE..81d1137D}. 

Given the slow progress on this problem over the last several decades, why does it 
make sense to insist that  a variational argument for non-equilibrium systems ought to 
exist?  The question is multi-faceted, but as  we are working within the  context of 
general relativity let us seek inspiration from that theory. One of the most topical 
problems in gravitational physics involves  two stars (or black holes) in a binary system, 
that lose orbital energy through the emission of gravitational waves. The data for 
the celebrated binary pulsar PSR1913+16 (and several similar systems) demonstrate 
that this phenomenon is described to excellent precision by Einstein's theory. 
Gravitational-wave emission is obviously a dissipative mechanism, yet the underlying 
theory is obtained from an action. This example tells us that you can, indeed, 
use a variational strategy for dissipative problems (a similar argument was recently 
made in \cite{2013PhRvL.110q4301G}). The key insight is that \emph{all the energy in 
the system must be accounted for}.  In many ways this statement is trivial. If you account 
for all the energy in a given system, including the ``heat bath'', then there is no 
\emph{dissipation} as such. Rather, one would be trying to model the 
\emph{redistribution} of energy within the larger (now closed) system. Obviously, if the 
proposed binary system is alone in the Universe and the gravitational-wave emission is 
properly accounted for, then the system is conservative and there is no reason why the 
dynamics should not derive from an action. This may be an acceptable \emph{logical} 
argument, but how do you make it into a \emph{practical} proposition for a generic 
dissipative system? This is the question that motivates the present work. 

Building on recent efforts on the problem of heat in general relativity \cite{Lopez-Monsalvo11:_causal_heat,2011CQGra..28s5023A}, 
where progress was made by treating the system's entropy  as an additional field (at the 
hydrodynamics level), we aim to establish how the convective variational formulation for 
relativistic fluids \cite{taub54:_gr_variat_princ,1970PhRvD...2.2762S,carter89:_covar_theor_conduc,1991RSPSA.433...45C,1994IJMPD...3...15C,1994CQGra..11.2013C,1995NuPhB.454..402C,andersson07:_livrev} 
can be extended to account for dissipative mechanisms. A central issue in this 
development  concerns the second law of thermodynamics (which singles out the 
entropy as being ``special'' and which is intimately linked to any thermodynamical arrow 
of time argument). In the proposed approach, the functional form for the dissipative 
equations derives from the choice of action, but (just like in all other proposed 
formulations) the inequality associated with the second law is imposed by hand. This 
may seem like a trick --- sweeping the problem under the carpet --- and we would be the 
first to agree that the model remains incomplete at the fundamental physics level, but we 
nevertheless believe that our new approach paves the way for a better understanding of 
the link between physics on the small scale and macroscopic (fluid) dynamics.

It is useful to make clear under which conditions the model is intended to apply. Our 
analysis builds on a long tradition from the study of multi-component mixtures in 
chemistry \cite{onsager31:_symmetry,prigogine} and fluid dynamics \cite{drewpass}. 
That is, the focus is on systems where different constituents retain their identity as the 
system evolves. It is also worth noting that the final construction shares many features 
with extended irreversible thermodynamics \cite{joubook}. As we are working within the  
framework of fluid dynamics, the construction assumes that a system can be described 
as a number of distinct, not necessarily co-moving, ``fluids''. As discussed in 
\cite{2012PhRvD..86f3002H} this boils down to assuming that each constituent has a 
short enough internal length scale over which averaging can be carried out (this could 
be the mean free path associated with scattering off of particles of the same species, or 
the coherence length of a superfluid condensate), while any mechanism that  couples 
the flows acts on a larger scale (or a longer time scale). Archetypal systems of this kind 
are i) laboratory superfluids, where an inviscid condensate is weakly coupled to a 
``normal'' fluid consisting of thermal excitations (for  descriptions related to the present 
work, see \cite{carter94:_canon_formul_newton_superfl,2011IJMPD..20.1215A}), and ii) 
the coupled neutron superfluid/proton superconductor mixture in the outer core of a 
mature neutron star \cite{1991ApJ...380..515M,2011MNRAS.410..805G}. The model we 
discuss here does \underline{not} consider systems where one (or more) components 
are \underline{not} in the fluid regime. One can think of many such problems of interest, 
e.g. involving superfluids at low enough temperature that the thermal excitations are in 
the ballistic regime or  systems involving radiation. In principle,  the model can be 
extended to consider such cases but it is not our ambition to do so here.  

\subsection{A few comments on the state of the art}

The model described in this paper provides, in essence, an effective field theory for 
dissipative fluids in general relativity.  Yet, this model is quite distinct from other recent 
efforts that take quantum field theory as their starting point, e.g. the  bulk of the work on 
holographic fluid dynamics \cite{2009CQGra..26v4003R}. In order to appreciate the 
distinction between the different approaches, and understand the actual state-of-the-art 
in the general area of relativistic fluid dynamics,  one has to consider the literature at a 
level of detail that goes beyond the fluid equations of motion. The key difference stems 
from the tradition in two different application areas. The classical theory --- the setting for 
the present work --- aims to provide a fluid model suitable for applications in 
astrophysics and cosmology. A key aspect of this area is the connection with Einstein's 
theory of gravity and the role of the dynamical spacetime, e.g. the link between fluid 
dynamics and gravitational-wave emission. In contrast, holographic fluid dynamics is a 
high-profile area within high-energy physics, motivated by the celebrated AdS/CFT 
conjecture. A driving motivation for this programme is the fact that one can use weakly 
coupled (classical) gravity analogues to probe strongly coupled field theories. This is 
potentially very important since strongly coupled theories are not within reach of other 
methods, e.g. perturbation theory. The fact that such problems may be considered using 
classical gravity analogues (e.g. black-hole dynamics, which may be probed relatively 
straightforwardly) is clearly attractive. In terms of applications, the holography efforts 
have been (perhaps mainly) driven by particle physics and the need to describe the 
quark-gluon plasma in colliders like RHIC and the LHC (there has also been a recent 
drive towards problems in low-temperature physics). This situation is different from that 
in astrophysics in that the focus of particle physics tends to be on hot low-density matter, 
rather than the cold high-density matter relevant for neutron stars (say), and the 
spacetime in the holography approach can be considered flat/fixed without any loss of 
precision. 

One would, of course, expect the different approaches to be compatible at a more 
fundamental level. The final fluid equations should take the same form and one should 
be able to identify the ``same'' dissipative terms \cite{baier,2010CQGra..27b5006R}. 
However, the developments have not yet reached the stage where a comparison is 
non-trivial. Neither theory is  complete and there is no easy way to bridge between the 
two (there is no straightforward link between weakly and strongly coupled theories). The 
models have also been developed to different degrees of sophistication. The classical 
gravity approach has been extended to deal with complex systems involving different 
states of matter, like superfluidity and the entrainment effect (in neutron stars due to the 
strong interaction or Bragg scattering by the crust lattice) and is being applied in 
situations with direct relevance for observations. Meanwhile, the bulk of the holography 
models have been carried out for conformal fluids. This assumption is not relevant for 
``normal matter'', but  would apply at extremely high energies.  At a sufficiently high 
energy the thermal energy dominates and  you can ignore the scale associated with the 
mass (= chemical potential) of any particles involved. The problem then has only one 
scale (the temperature) and simplifies considerably. The assumption would be relevant 
for strongly coupled QCD (this is manifest in the MIT bag model which accounts for the 
difference in the quark masses as corrections to a model with conformal symmetry). 
Moreover, the conformal symmetry leads to the trace of the stress-energy tensor 
vanishing. This is a significant constraint on the theory.  More recently, there have  been 
efforts to move away from these restrictive assumptions, accounting for a finite chemical 
potential \cite{chem1,chem2} and additional degrees of freedom associated with 
superfluidity \cite{sfl1,sfl2} etcetera. The price you pay for making the hydrodynamics 
more complicated is increased complexity in the corresponding higher dimensional 
gravity problem (e.g. the inclusion of conserved charges that require a coupling to a 
gauge field, like in electromagnetism, or superfluidity which requires the consideration 
of black branes).  This obviously makes the analogue gravity problem more difficult, but 
it may still be easier to deal with than the strongly coupled field theory on the other side 
of the correspondence. Finally, it is  worth noting that the hydrodynamics obtained from 
the fluid-gravity correspondence is often in a fixed curved spacetime. In essence, this 
means that the description includes ``external forces''.  These will have to be removed 
before these models can be used as inspiration for physical systems like neutron stars 
and other relevant gravitational-wave sources.
 
\section{Convective variational multi-fluid systems}

Building on Carter's convective variational formulation \cite{carter89:_covar_theor_conduc,andersson07:_livrev}, 
there has been considerable recent progress on the modelling of multi-fluid systems in
general relativity.  In addition to the intrinsic elegance of an action principle, an
appealing feature of the variational approach is that once an ``equation of state'' for 
matter (which here takes on the role of the Lagrangian) is provided  the theory provides
the relation between the various currents and their conjugate momenta. Another key 
advantage of the variational derivation is that it is straightforward to incorporate 
additional fluid components \cite{andersson07:_livrev}. Hence, the extension to more 
complicated  systems is natural.

The variational discussion takes as its starting point the notion of local fluid elements. 
These elements must contain enough particles that well-defined averaged state 
parameters (pressure, temperature, and so on) exist and can be measured reliably (the 
response of the relevant ``device'' must be much faster than the local changes in the 
fluid due to statistical fluctuations). At the same time, the fluid elements must  be small 
enough that their respective number of particles is infinitesimal relative to the entire 
system. Finally, from the spacetime point-of-view the fluid elements should be 
particle-like in that they trace out distinct worldlines. In this description, a 
\emph{multi-fluid} system is such that several distinct components are able to flow more 
or less independently \cite{2012PhRvD..86f3002H}. That is, each ``fluid'' of the  system 
has its own set of worldlines that it follows without losing its ``chemical identity''.
The archetypal multi-fluid system is superfluid Helium, which is known to be well 
described by a two-fluid model  \cite{carter94:_canon_formul_newton_superfl,2011IJMPD..20.1215A}. 
The decoupling of the two components is due to the superfluidity which suppresses 
particle scattering and friction. Another, less obvious, setting involves the flow of heat. In 
that case, it has been shown that a model based on treating the entropy component as 
an additional ``fluid'' successfully resolves troublesome issues associated with causality 
and stability and also leads to the emergence of the expected second sound 
\cite{1991PhRvD..43.1223P,Andersson10:_causal_heat,Lopez-Monsalvo11:_causal_heat,2011CQGra..28s5023A}.

In the following, we will consider a system with $N_c$ independent constituents. Not all 
of these must flow independently. There are situations where it is important to keep track 
of the chemical composition of the various fluid elements, and a workable model must 
allow for this. Hence, we allow for the presence of $N_f$ ($\le N_c$) distinct flows, and 
associated fluxes $n_\x^a$, where the index $\x$ labels the components and $a$ is the 
spacetime index. The associated number density (as measured by a comoving 
observer) is given by $n_\x^2 = - g_{ab} n_\x^a n_\x^b$, where $g_{ab}$ is the 
spacetime metric (assumed to have signature +2 in the following), and the ``fluid 
particles'' associated with each flux have worldlines that follow from the unit four-velocity 
$u^a_\x = n^a_\x/n_\x$.  (Throughout the paper we work in geometric units where the 
speed of light is unity.) When $N_f = N_c$, each constituent can move independently of 
the others, but when $N_f < N_c$, some of the constituents  are locked. As an example, 
this would be the case in a non-zero temperature system with vanishing heat conduction 
where  the heat is advected with the flow (the matter and entropy have the same four-
velocity). In general, entropy is accounted for by treating it as a separate component 
(with zero rest mass).

For an isotropic system the matter Lagrangian, $\Lambda$, should be a relativistic 
invariant and hence depend only on covariant combinations of the fluxes. This  includes 
the relative flows between them;  one must consider both $n_\x^2$ and 
$n_{\x\y}^2 = - g_{ab}n_\x^a n_\y^b$,  with $\y \neq \x$. The latter encodes the so-called 
entrainment effect, which tilts the momenta with respect to the currents when two or 
more fluids are coupled \cite{andreev75:_three_velocity_hydro,andersson07:_livrev}. In 
the case of neutron stars, the strong interaction is known to induce entrainment between 
neutrons and protons in the star's core \cite{1984ApJ...282..533A}. Meanwhile, the 
entropy-matter entrainment has been shown to be a crucial feature of the multi-fluid 
approach to heat conduction  \cite{Andersson10:_causal_heat,Lopez-Monsalvo11:_causal_heat,2011CQGra..28s5023A}.

An arbitrary variation of $\Lambda$ with respect to the fluxes $n^a_\x$ and the metric 
gives (here and in the following we ignore terms that can be written as total derivatives, 
that is, we ignore ``surface terms'' in the action)
\beq
    \delta \left(\sqrt{- g} \Lambda\right) = \sqrt{- g} \left[\sum_{\x} 
    \mu^\x_a \delta n^a_\x + \frac{1}{2} \left(\Lambda g^{a b} + \sum_{\x} 
    n^a_\x \mu^b_\x\right) \delta g_{a b}\right] \ , \label{dlamb}
\eeq
where $g$ is the determinant of the metric and $\mu^\x_a$ are the individual momenta. 
These take the form 
\beq
     \mu^\X_{a} = g_{ab} \left(\B^{\X } n^b_\X + \sum_\y \A^{\X \Y} 
                   n^b_\Y\right) \ , 
                   \end{equation} 
with
\begin{equation}
\B^\x =  - 2 \frac{\partial \Lambda}{\partial 
                 n^2_{\X}}    \ , 
\eeq
 and
\beq
                  \A^{\X \Y} = \A^{\Y \X} = - \frac{\partial \Lambda}{\partial 
                 n^2_{\X \Y}} \quad , \quad \X \neq \Y \ . \label{coef12} 
\end{equation} 
Each momentum covector, $\mu^\X_a$, is  dynamically, and thermodynamically, 
conjugate to the respective number density current, $n^a_\X$, and the magnitude gives 
the chemical potential. The $\A^{\X \Y}$ coefficients quantify the entrainment between 
the $\x$ and $\y$ components.  

Equation \eqref{dlamb} illustrates why a variational derivation of fluid dynamics is 
nontrivial. As it stands, the variation of $\Lambda$ suggests that the equations of motion 
would be $\mu^\x_a=0$; in essence, none of the fluids carry  energy or momentum. This 
problem is resolved by imposing constraints on the fluxes. This can be done in different 
ways, but the route we promote here seems (at least to us) the most natural. 

\begin{figure} 
\centering
\includegraphics[height=10cm,clip]{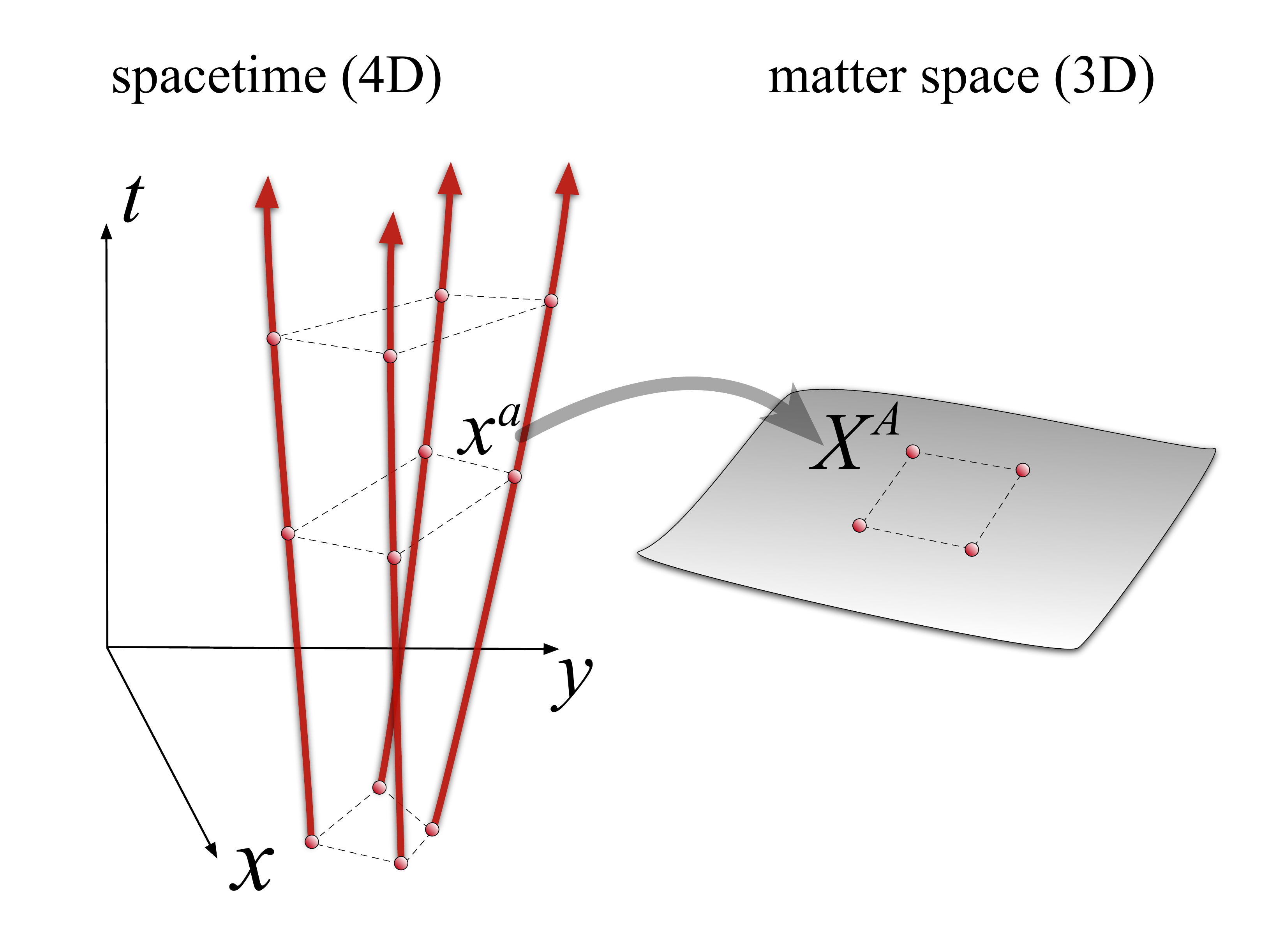}
\caption{An illustration of the pull-back formalism, where a given fluid is associated with 
a three-dimensional matter space. The coordinates of this space, $X^A$, essentially 
label the flowlines $x^a(\tau)$, where $\tau$ is a suitable parameter along each curve, 
of the various fluid elements in spacetime. These labels are assigned at the initial time 
of evolution, say $t=0$, and remain unchanged throughout.}
\label{matterspace}
\end{figure}

\begin{figure} 
\centering
\includegraphics[height=10cm,clip]{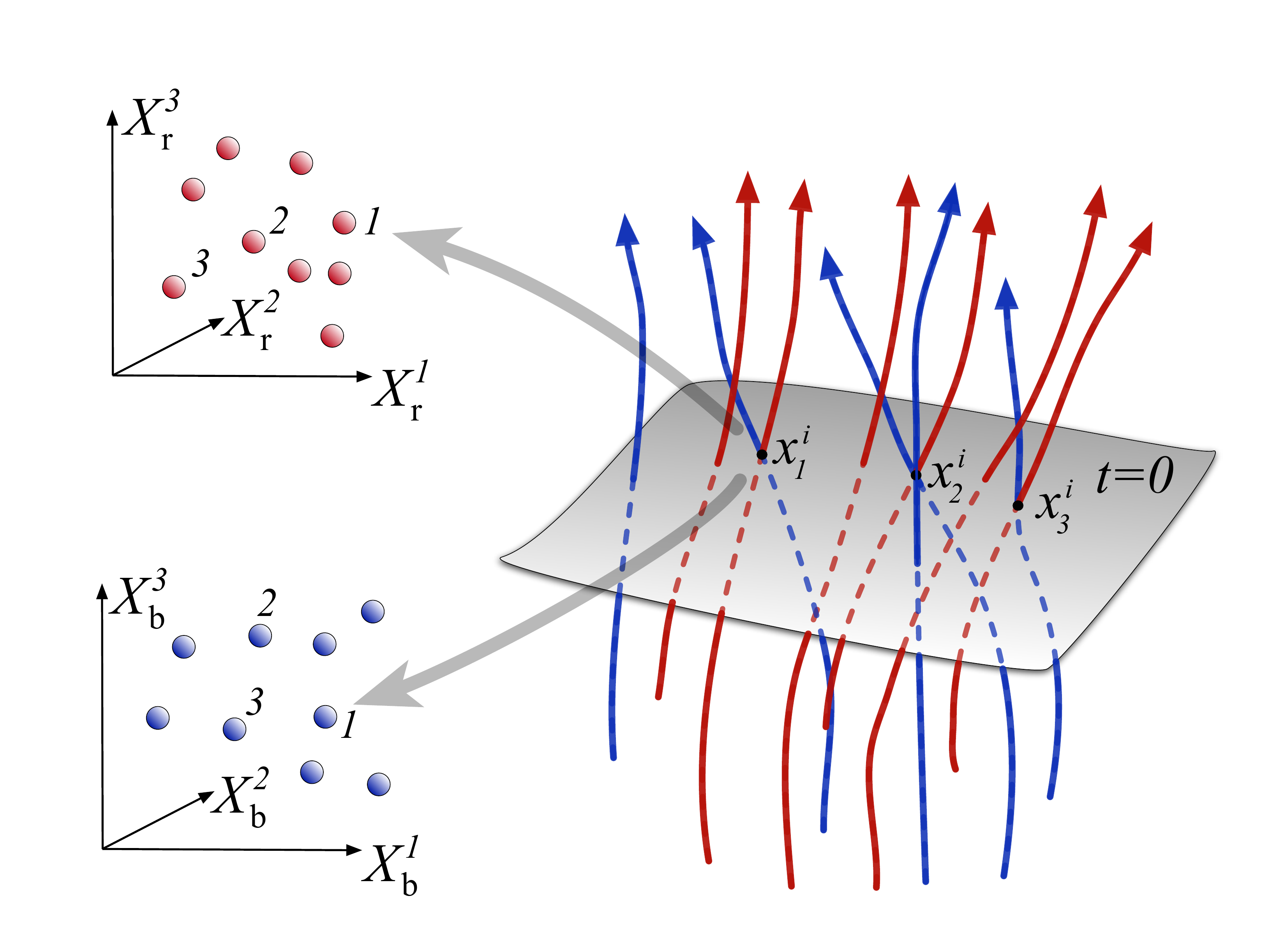}
\caption{In the case of systems with several coupled fluids  each component can be 
associated with its own three-dimensional matter space. The coordinates of this space, 
$X^A_\x$, which label the flowlines of the various fluid elements in spacetime,  are 
assigned at the initial time of evolution, say $t=0$. The illustration relates to a problem 
with two constituents, labelled x=r(ed) and x=b(lue). The map between each matter 
space and spacetime plays a key role in establishing the conservation of the matter 
flows in the variational model.}
\label{pullback}
\end{figure}

In fluid dynamics, there are two common approaches to monitoring the evolution: 
Eulerian and Lagrangian. In the former, an army of observers at rest with respect to a 
generic frame of reference make notes of the evolution as the various fluid elements 
intersect their worldlines. In the latter, each observer attaches him/herself to a particular 
fluid element and monitors how that element changes.  We take the Lagrangian 
point-of-view by introducing for each fluid an abstract three-dimensional ``matter'' space 
such that the worldline of a given fluid element is identified with a unique point in this 
space. The idea, which can be traced back to Taub \cite{taub54:_gr_variat_princ} (see 
also \cite{1970RSPSA.316....1O,1970JFM....44...19B,schutz77:_variat_aspec}), and 
which has featured prominently in the development of models for relativistic elasticity 
\cite{Grot1,Grot2,1972RSPSA.331...57C,carter73:_elast_pertur,1978JMP....19.1198M,1978JMP....19.1206M,1978JMP....19.1212M,1992JGP.....9..207K,2003CQGra..20..889B,1992GReGr..24..139M,2003CQGra..20.3613K,2012CQGra..29a5005G}, 
is illustrated in  Figure~\ref{matterspace}. The generalisation of the idea to the case 
were there are as many matter spaces as there are components is illustrated in 
Figure~\ref{pullback}. The coordinates of each matter space, $X^A_\x$ where 
$A = \{1,2,3\}$, serve  as  labels that distinguish fluid element worldlines. These labels 
are assigned at the initial time of the evolution, say $t=0$. The matter space coordinates 
can be considered as scalar fields on spacetime, with a unique map (obtained by a 
pull-back construction) relating them to the spacetime coordinates. We will demonstrate 
later that the $X_\x^A$ do not change along the associated  worldlines.

The variational construction involves three key steps. First we note that the conservation 
of the individual fluxes is ensured provided that the dual  three-form
\begin{equation}
 n^\X_{abc} = \epsilon_{abcd} n^{d}_\X   \quad , \quad
    n^a_\x = \frac{1}{3!} \epsilon^{a b c d} n^\x_{b c d} \ , \label{n3form}
\end{equation}
(where $\epsilon_{abcd}$ is the usual volume form associated with the spacetime) is 
closed;
\begin{equation} 
     \nabla_{[a} n^\X_{bcd]} = 0 
     \quad   \longrightarrow \quad 
  \nabla_{a} n^{a}_\X = 0  \label{consv2} 
\end{equation} 
(the square brackets indicate anti-symmetrization, as usual). In the second step we 
make use of the matter space to construct three-forms that are automatically closed on 
spacetime;
\beq
    n^\x_{a b c} =  \frac{\partial X^A_\x}{\partial x^{[a}} 
                   \frac{\partial X^B_\x}{\partial x^b} 
                   \frac{\partial X^C_\x}{\partial x^{c]}} 
                   n^\x_{A B C} \ , \label{pb3form}
\eeq 
where the Einstein summation convention applies to repeated matter space indices 
$A,B,C$. Here, and in the following, we use notation such that a spacetime object and 
its matter space image are represented by the same symbol, with only the indices being 
different (i.e. $n^\x_{abc}\leftrightarrow n^\x_{ABC}$). The volume form $n^\x_{ABC}$, 
which is assumed to be anti-symmetric, provides matter space with a geometric 
structure. If integrated over a volume in matter space it provides a measure of the 
number of particles in that volume.  

With the above definition, the three form \eqref{pb3form} is closed provided 
$n^\x_{A B C}$ is a function of the $X^A_\x$. In other words, if we take the scalar fields 
$X^A_\X$  to be the fundamental variables\footnote{It is worth pointing out that one can 
easily construct a variational model where the scalar fields $X^A_\x$ are the primary 
variables, satisfying the standard Euler-Lagrange equations (see 
\cite{comer93:_hamil_multi_con,comer94:_hamil_sf} for early work in this direction). 
This approach, recently explored in \cite{alt1,alt2,alt3,alt4,alt5}, is simply a reformulation 
of Carter's model which forms the basis of our work 
\cite{carter89:_covar_theor_conduc,andersson07:_livrev}.} they yield a representation 
for each particle number density current that is automatically conserved. Hence, it is 
natural to express the variations of the spacetime three-form in terms of the $X^A_\X$.

The final step involves introducing Lagrangian displacements $\xi^a_\X$  for each fluid. 
These displacements track the movement of the worldline of a given fluid element. From 
the standard definition of  Lagrangian variations in the relativistic context, see for 
example \cite{1975ApJ...200..204F,1978CMaPh..62..247F}, we have 
\beq
\Delta_\x X_\x^A = \delta X_\x^A + \mathcal{L}_{\xi_\x} X_\x^A = 0 \ , 
\label{DelX}
\eeq
where $\delta X^A_\X$ is the Eulerian variation and $\mathcal{L}_{\xi_\x}$ is the Lie 
derivative along $\xi_\x^a$. This means that convective variations are such that (since 
$ X_\x^A$ acts as a scalar field on spacetime)
\beq
    \delta X^A_\X = -  \mathcal{L}_{\xi_\x} X_\x^A = -  \xi^a_\x 
    \frac{\partial X^A_\x}{\partial x^a} \ . 
    \label{xlagfl}
\end{equation} 
After some algebra, one finds that this leads to
\beq
\Delta_\x n^\x_{abc} = 0 \ , 
\eeq
which in turn implies
\beq
\delta n_\x^a  = n_\x^b \nabla_b \xi^a_\x - \xi_\x^b 
                   \nabla_b n_\x^a - n^a_\x \left(\nabla_b 
                   \xi_\x^b + \frac{1}{2} g^{b c} \delta 
                   g_{b c}\right) \ . \label{delnvec} 
\end{equation} 
This is the result we require.  By expressing the variations of the matter Lagrangian in 
terms of the displacements $\xi^a_\x$ we ensure that the flux conservation is accounted 
for in the equations of motion. The variation of $\Lambda$ now leads to 
\beq
    \delta \left(\sqrt{- g} \Lambda\right) =  \sqrt{- g} \left[ 
   \frac{1}{2} \left(\Psi \delta^a{}_b + \sum_{\X } n^{a}_\X 
    \mu_b^\X\right) g^{b c} \delta g_{a c} -  
   \sum_{\X } f^\X_{a} \xi^{a}_\X\right]
    \ , \label{variable}
\eeq
where we have introduced the fluid force 
\beq 
     f^\X_b = n^a_\X \omega^\x_{a b} \ ,
\eeq
and the fluid vorticity
\beq
    \omega^\x_{a b} = 2 \nabla_{[a} \mu^\x_{b]} \ . \label{omx}
\eeq

From the constrained variation it thus follows that the equations of motion are simply 
given by\footnote{At this point we have  made a subtle switch: $f^\x_a = 0$ is enough if 
we still have in mind that the fluxes are functions of the $X^A_\x$, and those are what 
we solve for.  However, usually we have in mind that we are going to solve for the 
$n^a_\x$, in which case $ \nabla_a n^a_\x =0$ also has to be considered as an 
``equation of motion''. }
\beq
f^\x_a = 0  \quad \longrightarrow \quad 2 n^a_\X \nabla_{[a} \mu^\x_{b]} = 0 \ . 
\label{forcex}
\eeq

We also see that the stress-energy tensor (the variation with respect to the spacetime 
metric) takes the form
\begin{equation} 
     T^{a}{}_{b} = \Psi \delta^a{}_b + \sum_{\X} 
                       n^a_\X \mu^\X_b \ , 
\label{tab}\eeq
where
\beq
    \Psi = \Lambda - \sum_{\X} n^{a}_\X \mu^\X_{a} \ ,   \label{seten2} 
\end{equation} 
is the  (generalized) pressure. When the  set of  equations \eqref{forcex} are satisfied 
then it is automatically true that $\nabla_{a} T^{a}{}_{b} = 0$.

Over the last decade or so, the variational model has been applied to a range of 
interesting and relevant problems. This has led to progress in a number of  directions. 
Some of the results have been conceptual while others relate directly to applications. 
Briefly summarised;
\begin{itemize}

\item[i)] The variational model provides a natural framework to describe superfluid 
systems, both in the laboratory context  and in astrophysics. The associated quantization 
of vorticity is easily imposed on the canonical momentum, and the implications for the 
dynamics become quite intuitive. In the case of neutron star modelling, the entrainment 
plays an important role \cite{1984ApJ...282..533A}, so the fact that it is naturally included 
in the model is a great advantage.

\item[ii)] Since the variational construction makes direct use of  Lagrangian 
displacements and the matter space, it is straightforward to include the effects of 
elasticity in the formalism \cite{1972RSPSA.331...57C,carter73:_elast_pertur,2003CQGra..20..889B,1992GReGr..24..139M,2003CQGra..20.3613K,2012CQGra..29a5005G}. 
At the linear level, this simply amounts to keeping track of the deviations away from a 
relaxed reference configuration for which the strain vanishes. This has allowed realistic 
modelling of the dynamics of neutron star crusts 
\cite{carter06:_crust,Samuelsson07:_axial_crust,2009CQGra..26o5016S}. 

\item[iii)] The model has allowed us to make progress on the long-standing problem of 
heat-flux in general relativity \cite{carter83:_in_random_walk,carter_heat,1989PhLA..141..125H,1990PhRvD..41.3687O}, 
resolving issues regarding causality and stability \cite{hiscock85:_rel_diss_fluids,1990PhRvD..41.3687O}. 
Identifying one of the fluid components as the entropy (appropriate when the ``phonons'' 
in the system have a short enough mean-free path) and introducing a 
phenomenological ``resistivity'' one readily arrives at a formulation that honours the 
second law of thermodynamics and  exhibits the anticipated second sound for heat in 
the relevant limit \cite{2011CQGra..28s5023A}. The entropy entrainment provides a key 
ingredient in the model, encoding the inertia of heat which is required to ensure 
causality  \cite{Andersson10:_causal_heat,Lopez-Monsalvo11:_causal_heat}. 

\item[iv)] Due to its variational origin, it is relatively easy to extend the model to account 
for charged components and electromagnetism (via the standard gauge-coupling) 
\cite{carter80rheo}. In this case, the introduction of a phenomenological resistivity leads 
to a consistent derivation of the relativistic Ohm's law for two-components plasmas 
\cite{2012PhRvD..86d3002A}. The model can also be extended to account for finite 
temperature effects and the route to more complex models is (at least conceptually) 
quite clear.
\end{itemize}

\section{A new strategy for dissipative systems}

As we have already mentioned, the notion that one cannot use a variational approach to 
model dissipative systems seems somewhat at odds with the tenets of general relativity.
Einstein's field equations can be obtained from a variational principle, and if matter is 
included in the model then the stress-energy tensor follows  (at least in principle) from a 
variation with respect to the metric. There is no reason why this argument should not 
remain valid also for dissipative systems. As long as all energy contributions (matter, 
entropy, etcetera) are included the system is, in fact, ``closed'' and should lend itself to a 
variational analysis. Our aim is to develop this strategy in detail (in a way that differs 
substantially from Carter's approach in \cite{1991RSPSA.433...45C}). Ultimately, we are 
hoping to develop a practical model for dissipative multi-fluid dynamics which can be 
applied to a wide range of topical problems.  

We will now  take the first few steps towards this goal  by devising a variational 
argument that leads to the functional form of the dissipative fluid equations. The relevant 
dissipation coefficients are, in principle, calculable within the model although this would 
require a specific equation of state (in the form of an energy functional) to be provided. 
We do not address that problem here, preferring to focus on the formal construction of 
the variational model. In many ways, this is the same attitude as in classical mechanics 
where the equations of motion for a system can be written down without actual reference 
to a particular form for the energy. The completion of the model is, of course, important 
but the problem is sufficiently complex that it is sensible to progress in manageable 
steps. Moreover, we will demonstrate that we can make progress without considering a 
specific problem setting.  

\subsection{Interacting matter spaces}

The idea behind the new approach is, conceptually, quite simple.  Recalling that the 
individual matter spaces (associated with the various fluid components) play a central 
role in the variational construction for a conservative system, let us consider the 
``physics'' of a dissipative system, e.g. with resistivity, shear or bulk viscosity etcetera. 
On the micro-scale dissipation arises due to particle interactions/reactions. On the fluid 
scale this naturally translates into an \emph{interaction between the matter spaces}. As 
we will  demonstrate, this can be accounted for by letting each matter space be 
endowed with a volume form which depends on:
\begin{enumerate}
\item the coordinates of {\em all} the matter spaces, and 
\item the independent mappings of the spacetime metric into these spaces. 
\end{enumerate}
For example, if each $n^\x_{A B C}$ is no longer just a function of its own $X^A_\x$, the 
closure of $n^\x_{a b c}$ will be broken. \ As the fluxes are no longer  conserved, the 
formalism incorporates dissipation.

To see how this could work, let us revisit the conservative problem. Recall that the scalar 
fields $X^A_\x$ label the (fluid) particles. If these are conserved, then the $X^A_\x$ must 
be constant along the relevant worldlines. That this is, indeed, the case is easy to 
demonstrate. Letting $\tau_\x$ be the proper time of each worldline, we have
\beq
             \frac{{\rm d} X^A_\x}{{\rm d} \tau_\x} = u^a_\x 
             \frac{\partial X^A_\x}{\partial x^a} = \frac{1}{n_\x} n^\x_{B C D} \epsilon^{a b c d} 
             \frac{\partial X^A_\x}{\partial x^{a}}
             \frac{\partial X^B_\x}{\partial x^b} 
             \frac{\partial X^C_\x}{\partial x^c} 
             \frac{\partial X^D_\x}{\partial x^{d}} =0  \ . \label{xAdrag}                   
\eeq
Since a fluid element's matter space coordinates $X^A_\x$ are constant along its 
worldline, it must also be the case that  
\beq
\frac{{\rm d} n^\x_{A B C}}{{\rm d} \tau_\x }= 0 \ .
\eeq  
In other words, the volume form $n^\x_{ABC}$ is fixed in the associated matter space.

\begin{figure} 
\centering
\includegraphics[height=10cm,clip]{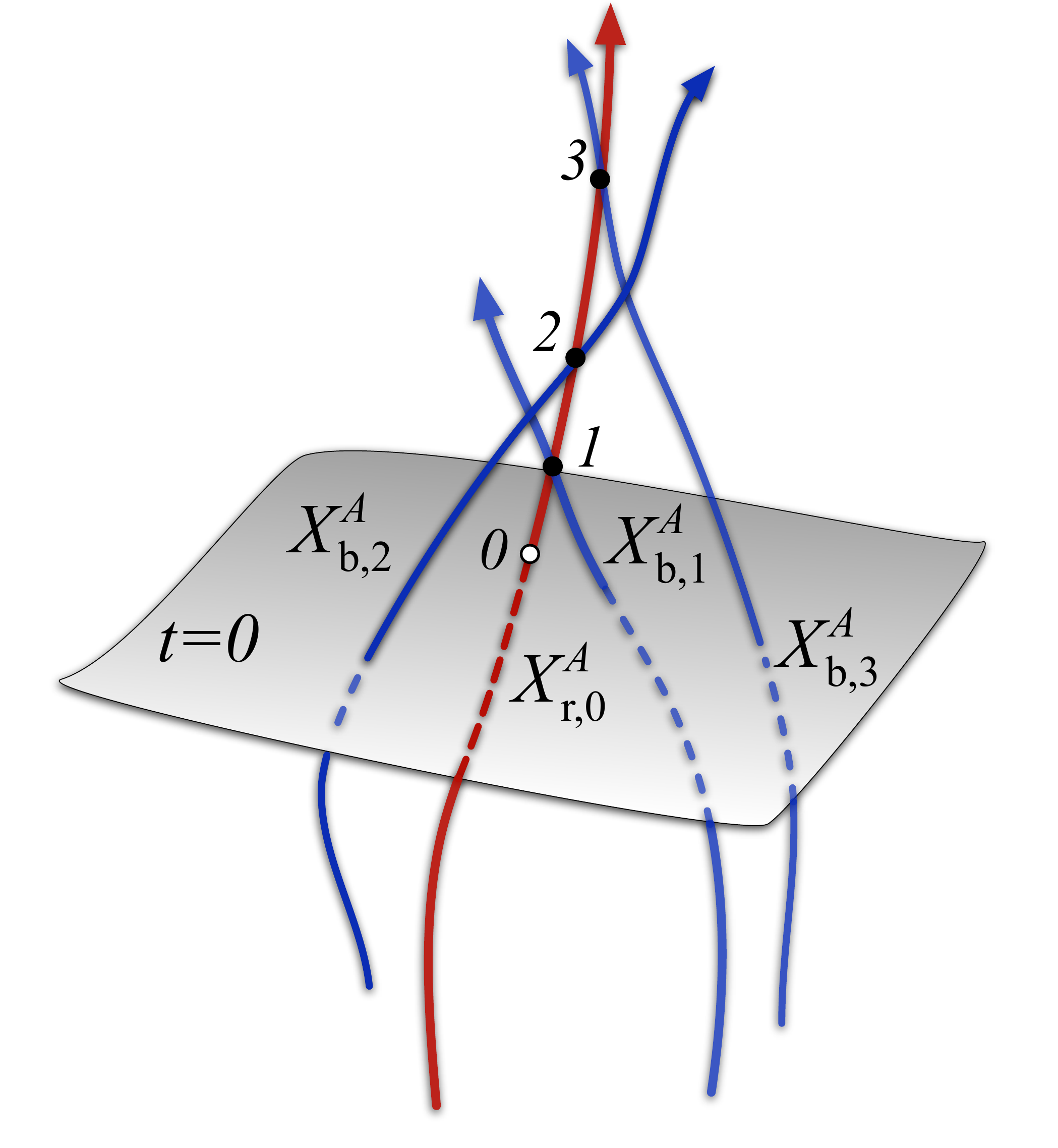}
\caption{An illustration of the notion that a coupling between  matter spaces can lead to 
dissipation. We consider the case of two fluids, labelled r and b (for red and blue). The 
individual $X^A_\x$ (assigned at the initial time, $t=0$) do not vary along their own 
worldlines, even when the system is dissipative. By adding $X^A_\y$ ($\y\neq\x$) we 
get ``evolution'' since  the worldlines cut across each other. Let us choose a particular 
worldline of the r-fluid, say $X^A_\mathrm{r,0}$, meaning that $X^A_\mathrm{r}$ will 
take the same value at each spacetime point $x^a$along the worldline. \ At an 
intersection with a worldline of a fluid element of the b-fluid (the point labelled 1 in the 
figure, say) the other fluid's worldline will have its own label (in this case 
$X^A_\mathrm{b,1}$), which is the same  at every point on that worldline. At the next 
intersection (point 2), the worldline we are following has the same value for 
$X^A_\mathrm{r}$, but it is intersected by a different worldline from the other fluid 
($X^A_\mathrm{b,2}$), meaning that $X^A_\mathrm{b}$ at each intersection is different. 
Hence, $X^A_\mathrm{b}$, when considered as a field in spacetime, must vary along 
the r-fluid worldlines, and vice versa. This is how the closure of the individual volume 
three-forms is broken and ultimately why the model is dissipative.}
\label{timeflow}
\end{figure}

It is clear from the steps required in this demonstration that the key to non-conservation 
is to allow  $n^\x_{A B C}$ to be a function of more than the $X^A_\x$. This is quite 
intuitive. The worldlines of the various fluids will in general cut across each other, 
leading to interactions/reactions. A more general functional form for the matter space 
volume forms $n_{ABC}^\x$ may be used to reflect this aspect of the underlying physics. 
A schematic illustration of how this works is provided in Figure~\ref{timeflow}.

As we will demonstrate in the next few sections, the simple step of enlarging the 
functional dependence of $n^\x_{ABC}$ does indeed allow us to build a variational 
model that incorporates the ``expected'' dissipative terms. However, it also takes us into 
territory where one has to tread carefully. In particular, one must pay more attention to 
the various ``matter space objects''.  We are now dealing with geometric objects that 
actually live in the higher-dimensional combination of \underline{all} the matter spaces, 
e.g. we are dealing with an object of the form
\beq
n^\x_{ABC} \left( X_\x^D, X_\y^E \right) dX_\x^A \wedge dX_\x^B \wedge dX_\x^C \ , 
\qquad \y \neq \x \ .
\eeq
That is, a volume form in the x-matter space parameterised by points in the y-matter 
spaces. From a presentational point-of-view we can still pretend that the individual 
matter spaces (related to spacetime via the same maps as in the conserved case) 
remain somehow ``distinct'', but in reality this is not the case. This issue requires more 
detailed analysis in the future.

The new model thus involves a change of emphasis. In the conservative multi-fluid 
problem one may, once the constrained variation is devised,  consider the various fluxes 
as the main variables and formulate the problem at the space-time level without 
bothering too much with the detailed matter space quantities. In the model  we advocate 
here, this is no longer the case. The change is inspired by efforts to model relativistic elasticity \cite{1972RSPSA.331...57C,carter73:_elast_pertur,2003CQGra..20..889B,1992GReGr..24..139M,2003CQGra..20.3613K,2012CQGra..29a5005G}, 
where  the role of the matter space is elevated and  the action is constructed at that 
level. In the case of elasticity, the fact that $n^\x_{ABC}$ is a fixed tensor in matter space 
allows the introduction of an associated ``matter space metric'' which can be used to 
quantify the deviation from the relaxed reference shape and hence account for elastic 
properties. 

When we allow $n^\x_{ABC}$ to be more complex we (inevitably) break some of the 
attractive features of the conservative model. Obviously, $n^\x_{ABC}$ is no longer a 
fixed matter space object. This has a number of repercussions, especially for 
discussions of elastic matter. We will not discuss those here, although it is worth noting 
that it is a very interesting problem given the obvious connection with visco-elasticity. 
Instead, we simply note that we can still construct the action from matter space objects. 
To do this we need the map of the spacetime metric into the relevant matter space: 
\beq
    \hxab = \frac{\partial X^A_\x}{\partial x^a}  
                  \frac{\partial X^B_\x}{\partial x^b} g^{a b} 
              = \hxba \ . \label{gmap}
\eeq
Note that  $\hxab$ is not likely to be a tensor on matter space. In order for that to be the 
case, the corresponding spacetime tensor must satisfy two conditions: First, it must be 
flowline orthogonal (on each index).  This holds for the present problem, since the 
operator which generates projections orthogonal to $\x$-fluid worldlines is  
\beq
\perp_\x^{ab} = g^{ab} + u_\x^a u_\x^b \ ,
\eeq
and because of Eq.~\eqref{xAdrag} we have
\beq
    \hxab = \frac{\partial X^A_\x}{\partial x^a}  
                  \frac{\partial X^B_\x}{\partial x^b} g^{a b} = \frac{\partial X^A_\x}{\partial x^a}  
                  \frac{\partial X^B_\x}{\partial x^b} \perp_\x^{a b} \ .
\eeq
The second condition that $\perp^{ab}_\x$ must satisfy so that $\hxab$ is a matter space 
tensor is \cite{2003CQGra..20..889B}
\beq
\mathcal{L}_{u_\x} \perp^{ab}_\x = 0 \ .
\eeq
This is not the case here; indeed, it is too severe for most relevant applications.

Anyway,  it is easy to show that a scalar constructed from the contraction involving 
$g^{ab}$ and some tensor $t^\x_{a\ldots}$ is identical to the analogous contraction of 
the corresponding matter space objects \cite{2003CQGra..20.3613K}. In particular, the 
number density follows from
\beq
n_\x^2 = - g_{ab} n_\x^a n_\x^b = \frac{1}{3!} g^{ad} g^{be} g^{cf} n^\x_{abc} n^\x_{def} =  
\frac{1}{3!} g_\x^{AD} g_\x^{BE} g^{CF} n^\x_{ABC} n^\x_{DEF} \ , 
\eeq
while the chemical potential 
\beq
        \mu^\x = - u_\x^a \mu^\x_a 
\eeq
(according to an observer at rest in the respective fluid's frame) can be obtained from
\beq
n_\x \mu^\x = - n_\x^a \mu^\x_a = \frac{1}{3!} \mu_\x^{abc} n^\x_{abc} =  \frac{1}{3!} 
\mu_\x^{ABC} n^\x_{ABC} \ . 
\eeq
Here we have introduced the dual to the momentum $\mu^\x_a$;
\beq
    \mu_\x^{a b c} = \epsilon^{a b c d} \mu_d^\x 
                     \quad , \quad
    \mu_a^\x = \frac{1}{3!} \epsilon_{a b c d} \mu_\x^{b c d} 
               \ , \label{mu3form}
\eeq
and its matter space image;
\beq
     \mu^{A B C}_\x = \frac{\partial X^A_\x}{\partial x^{[a}} 
                                \frac{\partial X^B_\x}{\partial x^b} 
                                \frac{\partial X^C_\x}{\partial x^{c]}}  \mu_\x^{a b c} \ .
\eeq 

The key take-home message is that we can think of the matter  action as being 
constructed entirely from matter space quantities. In the simplest case of a single 
component one would have $\Lambda \left( n_\x\right) = \Lambda \left( n^\x_{abc}, 
g^{ab}\right) \leftrightarrow \Lambda\left( n^\x_{ABC}, g_\x^{AB} \right)$. The 
specification of such an equation of state, with the functional dependencies discussed 
later, will eventually be required in order to complete the model we are designing. For 
the moment, we assume that this problem can be dealt with and move on to the actual 
variational equations of motion.

\subsection{Proof-of-principle: A reactive/resistive system}

As a first step towards making the proposal concrete, let us work through the key steps 
in the variational analysis, this time allowing for general variations of the matter space 
density. The matter space coordinates  still vary according to \eqref{xlagfl} (this is 
essentially just the definition of the Lagrangian displacement). Noting that 
\beq
\Delta_\x \left( \frac{\partial X_\x^A}{\partial x^a}\right) = \frac{\partial}{\partial x^a} 
\left( \Delta_\x X_\x^A \right) = 0 \ , 
\eeq
we easily arrive at the generic variation 
\beq
   \delta n^\x_{a b c} = - \lefx n^\x_{a b c} + 
                         \frac{\partial X^A_\x}{\partial x^{[a}} 
                         \frac{\partial X^B_\x}{\partial x^b} 
                         \frac{\partial X^C_\x}{\partial x^{c]}} 
                        \Delta_\x n^\x_{A B C} \ . \label{nxvargen}
\eeq
To make contact with \eqref{delnvec} we need 
\beq
\mu^\x_a \delta n^a_\x = \frac{1}{3!} \mu^\x_a \delta\left(  \epsilon^{abcd}
  n^\x_{bcd} \right) = - \frac{1}{3!} \mu_\x^{bcd} \delta n^\x_{bcd} + \frac{1}{3!} \mu^\x_a 
  n^\x_{bcd} \delta \epsilon^{abcd} \ ,
\eeq
and \cite{andersson07:_livrev}
\beq
\delta \epsilon^{abcd} = - \frac{1}{2} \epsilon^{abcd} g^{ef} \delta g _{ef}  \ .
\eeq
Hence,  we arrive at 
\beq
    \mu^\x_a \delta n^a_\x =  \frac{1}{3!} \mu^{a b c}_\x \lefx n^\x_{a b c} -  \frac{1}{2} 
    \mu^\x_a n^a_\x   g^{bc} \delta g_{bc} - 
     \frac{1}{3!} \mu^{A B C}_\x \Delta_\x n^\x_{A B C}  \ ,  \label{munxvar}
\eeq                 
and the ``final'' expression:                     
\beq
      \mu^\x_a \delta n^a_\x = \mu^\x_a \left(  n^b_\x \nabla_b \fd^a_\x - 
              \fd^b_\x  \nabla_b n^a_\x -  n^a_\x \nabla_b \fd^b_\x 
             - \frac{1}{2} n^a_\x g^{bc} \delta g_{bc}    \right)  - 
                     \frac{1}{3!} \mu^{A B C}_\x \Delta_\x n^\x_{A B C}   \ .
              \label{mudgen} 
\eeq
The terms in the bracket are the same as in the conservative case, cf. \eqref{delnvec}. 
The last term is new. 

The functional dependence of the volume form for a given fluid's matter space
is the main input for what follows. Obviously,  $n^\x_{A B C}$ must depend on $X^A_\x$, 
the coordinates of the corresponding matter space. This leads to the conservative 
dynamics. Adding to this,  let us include the coordinates $X^A_\y$ from the other, $\y 
\neq \x$, matter spaces. As we have already seen, this breaks the closure of 
$n^\x_{a b c}$. 

The required variation of $n^\x_{A B C}$ is now [in view of \eqref{DelX}]
\beq
  \Delta_\x n^\x_{A B C} = \sum_{\y \neq \x} 
         \frac{\partial n^\x_{A B C}}{\partial X^D_\y} \Delta_\x X^D_\y  = \sum_{\y \neq \x} 
         \frac{\partial n^\x_{A B C}}{\partial X^D_\y} \left(\xi_\x^a-\xi_\y^a\right) 
         \partial_a X^D_\y\ .
\label{DelNx}
\eeq
Comparing to \eqref{munxvar}, we see that it is natural to define
\beq
  \RXYa \equiv  \frac{1}{3!} \mu^{ABC}_\x  
           \frac{\partial n^\x_{ABC}}{\partial X^D_\y}  \partial_a X^D_\y \ . 
         \label{resist}   
\eeq
We then have
\beq
      \mu^\x_a \delta n^a_\x = \mu^\x_a \left(  n^b_\x \nabla_b \fd^a_\x - 
              \fd^b_\x  \nabla_b n^a_\x -  n^a_\x \nabla_b \fd^b_\x 
             - \frac{1}{2} n^a_\x g^{bc} \delta g_{bc}    \right)  +  \sum_{\y \neq \x} \RXYa 
              \left(\fd_\y^a - \fd_\x^a \right)
                \ .
              \label{mudnxgen} 
\eeq

The final step of the exercise involves writing down the variation of the matter 
Lagrangian, $\Lambda$. Starting from \eqref{dlamb}, we arrive at 
\beq
    \delta \left(\sqrt{- g} \Lambda\right) = - \sqrt{- g} \left\{\sum_{\x} \left( f^\x_a +\mu^\x_a 
    \Gamma_\x  -  R^{\x}_a\right)  \fd^a_\x - \frac{1}{2} \left(\Psi  g^{a b} + \sum_{\x} n^a_\x 
    \mu^b_\x  \right) \delta g_{a b}\right\}  \ , \label{varladiss}
\eeq
where we have used 
\beq
  \sum_{\x} \sum_{\y \neq \x} \RXYa \fd^a_\y = 
  \sum_{\x} \sum_{\y \neq \x} \RYXa \fd^a_\x \ .
\label{RXYrel}\eeq
We have also defined
\beq
R^{\x}_a = \sum_{\y\neq\x} \left(  R^{\y\x}_a - R^{\x\y}_a \right) \ ,
\label{Rx}
\eeq
and
\beq
       \Gamma_\x = \nabla_a n^a_\x \ . \label{dualparc}
\eeq
Hence, the individual components are governed by the equations of motion
\beq
f^\x_a + \Gamma_\x \mu^\x_a  =  n^b_\x \omega^\x_{b a} +  \Gamma_\x 
\mu^\x_a = R^{\x}_a \ .
\label{force1}
\eeq
Since the force term $f^\x_a$ on the left-hand side is orthogonal to $n_\x^a$ (by the 
anti-symmetry of $\omega^\x_{a b}$) it is easy to see that this result implies that the 
particle creation/destruction rates are given by
\beq
\Gamma_\x = -  \frac{1}{\mu^\x} u_\x^a R^{\x}_a \ .
\eeq

Finally, an orthogonal projection of \eqref{force1} leads to
\beq
 2 n^a_\X \nabla_{[a} \mu^\x_{b]} 
+\Gamma_\x \perp_{\x b}^a \mu^\x_a  = \perp_{\x b}^a R^{\x}_a \ .
\eeq
These equations provide the dissipative equations of motion for this system.

With Equation~(\ref{varladiss}) we have a true action principle---in the sense that the 
field equations are extrema of the action---for a system of fluids that includes 
dissipation. In many ways, this demonstration is the key result of this work.

Before we move on, it is worth noting that the stress-energy tensor is still given by 
\eqref{tab} and we can show that 
\beq 
      \nabla_b T^b{}_a =  \sum_{\x}\left(  f^\x_a + \mu_a^\x \Gamma_\x \right) = 0 \ ,
\eeq
because
\beq
   \sum_{\x}\rtotxa = 0 \ , \label{resvan}
\eeq
identically. The requirement that the covariant divergence of the stress-energy tensor 
vanish is automatically guaranteed by the dissipative fluid equations, in keeping with the 
diffeomorphism invariance of the theory.

\subsection{The problem of heat revisited}

In much of the relevant literature, dissipative terms have been added to the equations of 
relativistic fluid dynamics in a somewhat ad hoc manner, inspired  by some level of 
intuition of how the system ``ought to behave'' (for recent examples, see \cite{2007PhRvC..76a4910M,2007PhRvC..76a4909M,2008PhRvC..77b4910P,2010CQGra..27b5006R,2010IJMPE..19....1R,2011arXiv1105.3733B,2011JPhCS.270a2042M,2012EPJA...48..166E,2013JHEP...02..153B,2013PhRvC..87b1901J}). 
The model developed in the previous section puts us in a rather different position as
the dissipative contributions were derived, not postulated. This leads to a number of 
interesting (and challenging) questions, most of which we are not in a position to answer 
at this point. It is, however, imperative that we establish that the construction ``makes 
sense''. To do this, we need to understand the physics content of the model. 

In order to gain insight, let us consider the simplest relevant setting. Assume that we 
consider a system with two components; matter (labelled $\n$) and heat, represented by 
the entropy (labelled $\s$).  In principle, we need to provide an equation of state (that 
satisfies relevant physics constraints) in order to complete the model. Once this is 
provided we can calculate the resistivity coefficients from \eqref{resist} and then model 
the system using the momentum equations \eqref{force1}. However, a discussion of 
suitable equations of state would force our attention away from the main focus here, the 
variational model itself. Hence, we prefer to consider the problem from a 
phenomenological point-of-view. This is sufficient if our main aim is to show that the 
model has the anticipated features. To make the model specific, let us assume that the 
matter component is conserved, but the entropy does not need to be. This is the problem 
of relativistic heat flow. This problem was recently considered in 
\cite{Lopez-Monsalvo11:_causal_heat,2011CQGra..28s5023A}, and it is useful to 
compare the present model to the results of that analysis. This problem is simple 
enough that we should be able to understand what is going on.

First of all, given that we only have two components then
\beq
R^{\n}_a = R^{\s\n}_a - R^{\n\s}_a = - R^\s_a \ .
\eeq
Secondly, the conservation of the material component implies that 
\beq
\Gamma_\n = - \frac{1}{\mu^\n } u_\n^a R^{\n}_a =  \frac{1}{\mu^\n } u_\n^a R^{\n\s}_a 
= 0 \quad \longrightarrow \quad   u_\n^a R^{\n\s}_a = 0 \ . 
\label{Gn}
\eeq
The upshot is that $R^{\n\s}_a$ must be orthogonal to \underline{both} $u_\n^a$ and 
$u_\s^a$. Meanwhile, the entropy change is constrained by the second law. That is, 
\beq
\Gamma_\s = -  \frac{1}{T} u_\s^a R^{\s}_a =  \frac{1}{T} u_\s^a R^{\s\n}_a \ge 0  \ , 
\label{Gs}
\eeq
where we have introduced the temperature $T = \mu^\s$. Note that the constraints affect 
the two, likely independent, contributions to $R^\n_a$. We cannot infer a link between 
$R^{\n\s}_a$ and $R^{\s\n}_a$ at this point.

So far we have not introduced a privileged observer. This is in contrast to most previous 
work which takes this as starting point for the discussion. This means that a direct 
comparison with other results, such as those in \cite{2007PhRvC..76a4910M,2007PhRvC..76a4909M,2008PhRvC..77b4910P,2010CQGra..27b5006R,2010IJMPE..19....1R,2011arXiv1105.3733B,2011JPhCS.270a2042M,2012EPJA...48..166E,2013JHEP...02..153B,2013PhRvC..87b1901J}, 
require a bit of effort. In order to facilitate a comparison, let us follow 
\cite{Lopez-Monsalvo11:_causal_heat} and focus on an observer moving along with the 
matter flow (in the spirit of Eckart \cite{eckart40:_rel_diss_fluid}). Thus we have 
$u^a=u_\n^a$ and the relative flow required to express the entropy flux is defined such 
that 
\beq
u_\s^a = \gamma \left( u^a + w^a \right) \ , \quad \mbox{where} \quad u^a w_a = 0\ , 
\quad \mbox{and} \quad \gamma = \left( 1 - w^2 \right)^{-1/2}  \ . 
\label{uxdecomp}
\eeq
The relative velocity $w^a$ is aligned with the heat flux vector (as discussed in 
\cite{Lopez-Monsalvo11:_causal_heat}).

Given \eqref{Gn} and \eqref{Gs} it  makes sense to introduce the decompositions
\beq
R^{\n\s}_a = \epsilon_{abcd} \phi_\n^b u^c w^d \ , 
\eeq
and 
\beq
R^{\s\n}_a = R_w w_a +  \epsilon_{abcd} \phi_\s^b u^c w^d \ , 
\label{Rsn1}\eeq
where $\phi_\n^a$ and $\phi_\s^a$ are unspecified vector fields.
We then see that \eqref{Gs} leads to\footnote{It is worth noting that we are making the 
standard assumption that the second law must hold locally for each fluid element. It is by 
no means obvious that this has to be true. The question relates to the size of the system 
from a statistical perspective. However, without this assumption you will not progress 
beyond this point.}
\beq
T \Gamma_\s =  \gamma R_w  w^2  \ge 0  \quad \longrightarrow \quad R_w > 0 \ . 
\label{second1}
\eeq
Meanwhile, the two components $\phi_\n^a$ and $\phi_\s^a$ are \underline{not} 
constrained by the thermodynamics. This leaves a degree of arbitrariness in the 
model.\footnote{Note that this arbitrariness would be removed if we provided a suitable 
equation of state, in which case the coefficients could be obtained from the definition 
\eqref{resist}.} Should we be surprised by this? Not really. A similar issue was, in fact, 
discussed in \cite{Lopez-Monsalvo11:_causal_heat} where it was demonstrated that the 
variational model led to the presence of a number of terms in the heat equation that 
could not be constrained by the second law. It was also pointed out that the difference 
between the model advocated in \cite{Lopez-Monsalvo11:_causal_heat} and the 
second-order model of Israel and Stewart appeared at this level 
\cite{1991PhRvD..43.1223P}. It has not been established whether there are situations 
where these terms  have a notable effect on the dynamics.  We leave this as an 
interesting question for the future.

\subsection{Adding dissipative stresses}

We have demonstrated how dissipation can be included in the variational multi-fluid 
formalism.  This is an important step towards a deeper conceptual understanding of 
non-equilibrium systems in general relativity. Dissipative contributions 
that have previously been {\em postulated} can now be {\em derived} from underlying 
principles. Moreover, as the comparison with the problem of heat flow demonstrates, the 
variational model points to new aspects of the problem. However, the example we 
considered above only accounts for two particular non-equilibrium phenomena, particle 
non-conservation and resistivity. In order to convincingly argue that our model 
represents a conceptual breakthrough, we need to demonstrate that the action principle 
generates terms in the field equations of the tensorial form expected for a more general 
range of  processes. Thus, we turn to the issue of dissipative stresses. 

The obvious starting point for an extension of the variational strategy is to ask what other 
quantities the matter space volume form, $n^\x_{ABC}$, may depend on. The natural 
object to consider is the mapping of the spacetime metric, $g_{a b}$, into the respective 
matter spaces.  As we will now demonstrate, this leads to a system with dissipative 
shear stresses. 

In mapping the metric into the matter spaces we have in principle three independent 
possibilities. Let us first consider the most intuitive option, which involves allowing 
$n^\x_{ABC}$ to depend on $\hxab$, as defined in \eqref{gmap}.

Noting that Eq.~(\ref{xAdrag}) implies that the $X^A_\x$ will still be conserved 
along the associated flow, the variation of $n^\x_{A B C}$ is now such that
\beq
  \Delta_\x n^\x_{A B C} = \frac{\partial n^\x_{A B C}}{\partial \hxde}
         \Delta_\x \hxde       + \sum_{\y \neq \x}
         \frac{\partial n^\x_{A B C}}{\partial X^D_\y}  \Delta_\x X^D_\y  \ .
\label{dnnew}\eeq
The first term  in this expression is new, the second term is the same as in 
\eqref{DelNx} . The new term is easily worked out, following the steps from the simpler 
model. We find that
\beq 
      \Delta_\x \hxab =  \frac{\partial X^A_\x}{\partial x^a} 
                                      \frac{\partial X^B_\x}{\partial x^b} \Delta_\x g^{ab} =
                                      \frac{\partial X^A_\x}{\partial x^a} 
                                      \frac{\partial X^B_\x}{\partial x^b}
                                      \left[\delta g^{a b}- 2 \nabla^{(a} \fd^{b)}_\x \right] \ , 
\eeq
where we have used 
\beq
\Delta_\x g^{ab} = \delta g^{a b}- 2 \nabla^{(a} \fd^{b)}_\x \ , 
\eeq
(round brackets indicate symmetrization, as usual.)

As in the previous example, the variation of the matter Lagrangian involves 
$\mu^{A B C}_\x \Delta_\x n^\x_{A B C}$. The new contribution takes the form
\begin{multline}
 \frac{1}{3!} \mu^{A B C}_\x  \frac{\partial n^\x_{A B C}}{\partial \hxde}
         \Delta_\x \hxde =  \frac{1}{3!} \mu^{A B C}_\x  
         \frac{\partial n^\x_{A B C}}{\partial \hxde}
         \frac{\partial X^D_\x}{\partial x^a} \frac{\partial X^E_\x}{\partial x^b}
         \left[\delta g^{a b} - 2 \nabla^{(a} \fd^{b)}_\x \right]  \\
         = - \frac{1}{2} \DXab \left[ g^{ac} g^{bd} \delta g_{cd} + 2 \nabla^{(a} \fd^{b)}_\x\right]  
         = - \frac{1}{2}  \DXuab \delta g_{ab} - \DXab \nabla^b  \xi_\x^a \ , 
\end{multline}
where we have defined
\beq
\DXab = \frac{1}{3}  \mu^{ABC}_\x \frac{\partial n^\x_{ABC}}{\partial \hxde}  
            \frac{\partial X^D_\x}{\partial x^a} \frac{\partial X^E_\x}{\partial x^b} = \DXba \ , 
            \label{dxab} 
\eeq
such that 
\beq
u_\x^a \DXba= 0   \ . 
\eeq

Combining the results, we arrive at 
\begin{multline}
      \mu^\x_a \delta n^a_\x = \mu^\x_a \left(  n^b_\x \nabla_b \fd^a_\x - 
              \fd^b_\x  \nabla_b n^a_\x -  n^a_\x \nabla_b \fd^b_\x   \right) + \DXab \nabla^b  
              \xi_\x^a \\
             + \sum_{\y \neq \x} \RXYa 
              \left(\fd_\y^a - \fd_\x^a \right)
             + \frac{1}{2} \left[ \mu^\x_c n^c_\x g^{ab} + \DXuab\right] \delta g_{ab}  \ .
\end{multline}

Introducing the total dissipative stresses, in this case trivially setting
\beq
\dtotxab = \DXab \ , 
\eeq
we see that  Eq.~(\ref{varladiss}) becomes
\begin{multline}
    \delta \left(\sqrt{- g} \Lambda\right) = - \sqrt{- g} \left\{\sum_{\x} \left(f^\x_a 
    + \Gamma_\x \mu^\x_a + \nabla^b\dtotxba  - \rtotxa\right)  \fd^a_\x \right.  \\
   \left.- \frac{1}{2} \left[\Psi  g^{a b} + \sum_{\x} \left(n^a_\x \mu^b_\x +  
    \dtotxab\right)\right] \delta g_{a b}\right\}  \ , 
\end{multline}
where we have used  \eqref{RXYrel} and \eqref{Rx}  for the resistivity currents.

The  equations of motion now take the form
\beq
    f^\x_a + \Gamma_\x \mu^\x_a + \nabla^b \dtotxab =  \rtotxa 
             \ , \label{xfleqn}
\eeq
and the stress-energy tensor is 
\beq
       T^{a b} = \Psi  g^{a b} + \sum_{\x} \left(n^a_\x \mu^b_\x +  \dtotxabu\right) \ ,
                        \label{emstens}
\eeq
where the generalised pressure, $\Psi$, remains unchanged, cf. \eqref{seten2}.
As in the previous problem, it is quite easy to show that 
\beq 
      \nabla_b T^b{}_a =  \sum_{\x} \left(f^\x_a + \Gamma_\x \mu^\x_a + 
                                          \nabla^b  \dtotxab\right) = 0 \ ,
\eeq
since \eqref{resvan} still holds. 

Finally, we can extract the various creation/destruction rates. We first contract 
Eq.~(\ref{xfleqn}) with $u^a_\x$, noting that $u^a_\x f^\x_a = 0$ and 
$u^a_\x \nabla^b \dtotxab = - \dtotxab \nabla^b u^a_\x$, to find 
\beq
       \mu^\x \Gamma_\x = - \rtotxa u^a_\x - \dtotxab \nabla^b u^a_\x    \ .
       \label{flxproj}
\eeq
When $\x=\s$ this gives the entropy creation rate which should be constrained by the 
second law.

\subsection{Rediscovering Navier-Stokes}

Armed with the more general constraint \eqref{flxproj} for the dissipative terms, let us 
revisit the model problem from Section~IIIB.  In the spirit of that discussion, let us ask 
what  we can learn from the various constraints that follow from the derivation (ignoring 
the fact that the coefficients involved could, at least in principle, be calculated from 
\eqref{resist} and \eqref{dxab} if we provided a suitable equation of state). That is,  we 
consider a two-component system with a material component (n) and entropy/heat (s) 
with the added physics input that $\Gamma_\n=0$. As in the previous discussion of this 
problem we will use an observer moving along with the matter flow, such that 
$u^a = u_\n^a$ and $w^a$ represents the relative flow.

Let us first consider  the matter component. Since we know that $R^{\n\s}$ should be 
orthogonal to $u_\s^a$ we introduce the decomposition
\beq
R^{\n\s}_a = R_u \left( w^2 u_a + w_a \right) + \epsilon_{abcd} \phi_\n^b u^c w^d \ .
\label{resistme} \eeq
Then \eqref{flxproj} implies that
\beq
D^\n_{ab} \nabla^b u^a = - R^{\n\s}_a u^a = R_u w^2  \ . 
\label{rel1} 
\eeq
Now, there are two cases one may consider. In the general situation, when there is a 
distinct heat flow, we have $w^2>0$ which if we take $R_u >0$ implies that the left-hand 
side of \eqref{rel1} must be positive. To ensure that this is the case, we use the standard 
decomposition (with the same conventions as in \cite{2012PhRvD..86d3002A})
\beq
\nabla_a u^\x_b = \sigma^\x_{ab} +\varpi^\x_{ab} + u^\x_a \dot{u}^\x_b +\frac{1}{3} 
\theta^\x \perp^\x_{ab}\ , 
\label{deco}\eeq
where 
\beq
\sigma^\x_{ab} = D_{\langle a}u^\x_{b\rangle} \ , \qquad \mbox{with} \qquad D_a u^\x_b 
= \perp^\x_{ac} \perp^\x_{bd} \nabla^c u_\x^d \ , 
\eeq
where the angular brackets indicate symmetrization and trace removal,
\beq
\varpi^\x_{ab} = D_{[a} u^\x_{b]} \ , 
\eeq
\beq
\theta^\x = \nabla_a u_\x^a \ , 
\eeq
and 
\beq
\dot u^\x_a = u_\x^b \nabla_b u^\x_a \ . 
\eeq

With these definitions, each term in \eqref{deco} is orthogonal to $u_\x^b$. From the fact 
that $\DXab$ is symmetric and orthogonal to $u_\x^a$ it is easy to see that the condition 
inferred from  \eqref{rel1} is satisfied provided we have
\beq
D^\n_{ab} = \eta^\n \sigma^\n_{ab} + \zeta^\n \theta^\n \perp^\n_{ab}
\eeq
with $\eta^\n>0$ and $\zeta^\n>0$. We recognise this as the dissipative (shear- and 
bulk viscosity) stresses expected in the Navier-Stokes equations. Interestingly, the 
second law of thermodynamics was not engaged in the derivation of this result. 

If, on the other hand, there is no heat flux in the system, then $w^2=0$ and we must 
have
\beq
\sigma^\n_{ab} = 0 \ , \qquad  \theta^\n =0  \ .
\eeq
These are, of course, the expected conditions for an equilibrium system. 

Let us now turn to the entropy condition. Making use of the results from the heat 
example discussed in section~IIIB, noting that we can still use \eqref{Rsn1} for 
$R^{\s\n}_a$, we see that \eqref{flxproj} leads to
\beq
T \Gamma_\s = \gamma R_w w^2 -  D^\s_{ab} \nabla^b u_\s^a \ge 0 \ , 
\eeq
as required by the second law. This suggests that, in addition to $R_w>0$ from before, 
we should have 
\beq
D^\s_{ab} =- \eta^\s \sigma^\s_{ab} - \zeta^\s \theta^\s \perp^\s_{ab}
\eeq
with $\eta^\s>0$ and $\zeta^\s>0$.

This example is a little bit more ``confusing'' than the pure heat conduction case.  On the 
one hand, it is impressive that we can arrive at the expected form of the equations from 
this rather general analysis. On the other hand, it is frustrating that we cannot pin down, 
for example, the sign of the friction coefficient in \eqref{resistme}. To do this, we need to 
consider a particular physics set-up where \eqref{resist} can be worked out. We plan to 
consider this  problem in more detail later. A particular system worth considering is 
superfluid Helium, for which the presence of several bulk viscosity channels -- most 
likely related to to $\zeta^\n$ and $\zeta^\s$ in the present model -- are known to exist
\cite{khalatnikov,putterman}.
  
\subsection{Adding dissipative stresses: A more general case}

The previous example demonstrates the promise of the new variational model. By 
allowing each matter space volume form $n^\x_{ABC}$ to depend on the coordinates of 
the other matter spaces $X^A_\y$ ($\y\neq\x$) as well as the mapping of the spacetime 
metric $g^{AB}_\x$, we arrive at a model that allows for particle non-conservation, 
resistivity and dissipative stresses. This is a conceptual success, but we now face a new 
set of questions. For example, it seems legitimate  to ask whether the model we 
developed in Section~IIIC is the most general construction. It is relatively easy to see 
that it is not; we could have considered other mappings of the metric. This problem turns 
out to be relevant, because it leads to a demonstration that the \emph{individual} 
dissipative stress tensors need not be symmetric even though \emph{the sum of them} 
is. The relevance of such asymmetries, and their potential role in modelling neutron star 
superfluids, has already been discussed in \cite{andersson05:_flux_con,2011IJMPD..20.1215A,2012PhRvD..86f3002H}.

In mapping the metric into the matter spaces in Section~IIIC we only considered one of 
the  three independent possibilities. We may also;
\begin{enumerate}
\item
allow $n^\x_{ABC}$ to depend on $g^{AB}_\y$, the metric mapped into the other matter 
spaces, or 
\item
use the mixed mapping\footnote{As in the previous model, it is worth noting that 
$\hxyab$ and $g^{AB}_\y$ are not tensors in the matter space of the x component. 
In this case, the spacetime objects are not even (completely)  flowline orthogonal with 
respect to $u_\x^a$. This is obvious from the fact that 
$$
  \hxyab = \frac{\partial X^A_\x}{\partial x^a}  
                      \frac{\partial X^B_\y}{\partial x^b} \perp_\x^{ac} \perp_{\y\ c}^b \ .
$$
} 
\beq
      \hxyab = \frac{\partial X^A_\x}{\partial x^a}  
                      \frac{\partial X^B_\y}{\partial x^b} g^{a b} \
                      \quad , \quad
       \y \neq \x \ .
\eeq
\item 
consider $n^\x_{ABC}$ as a function of $g^{AB}_{\y\z}$ where y and z are different, but 
neither is equal to x. (We will not work this case out explicitly; the extension is relatively 
straightforward given the details below. )
\end{enumerate}
It is worth noting that the only symmetry in exchange of indices for $\hxyab$ is
\beq
    \hxyab = \hyxba \ .
\eeq
This implies that $\hxylbabrb$ may not vanish, which in turn suggests the presence of 
the asymmetric terms among the dissipative stresses.

The variation of $n^\x_{A B C}$ is now such that
\bea
  \Delta_\x n^\x_{A B C} &=& \frac{\partial n^\x_{A B C}}{\partial \hxde}
         \Delta_\x \hxde \cr
         &&+ \sum_{\y \neq \x} \left(
         \frac{\partial n^\x_{A B C}}{\partial X^D_\y}  \Delta_\x X^D_\y + 
         \frac{\partial n^\x_{A B C}}{\partial \hxyde}
         \Delta_\x \hxyde + \frac{\partial n^\x_{A B C}}{\partial \hyde}
        \Delta_\x \hyde\right) \ .
\eea
Comparing to \eqref{dnnew} we have two new terms;
\beq
     \Delta_\x \hxyab  = \frac{\partial X^A_\x}{\partial x^a} 
                                     \frac{\partial X^B_\y}{\partial x^b} \left[\delta g^{a b} - 2\nabla^{(a} 
                                     \fd^{b)}_\x  \right] + g^{a b} 
                                     \frac{\partial X^A_\x}{\partial x^a}  \left(\lefx - \lefy\right) 
                                     \frac{\partial X^B_\y}{\partial x^b} \ ,
\eeq
and
\beq
    \Delta_\x \hyab=  \frac{\partial X^A_\y}{\partial x^a} 
                                     \frac{\partial X^B_\y}{\partial x^b}
                                     \left[\delta g^{a b}- 2 \nabla^{(a} \fd^{b)}_\x \right] \\
                                     + g^{ab} \left(\lefx - \lefy\right) \left( \frac{\partial X_\y^A}{\partial x^a}  
                                     \frac{\partial X_\y^B}{\partial x^b}  \right)
                                     \ .
\eeq

In order to build the variation of the matter Lagrangian we need
\begin{multline}
 \frac{1}{3!}\mu^{A B C}_\x \frac{\partial n^\x_{A B C}}{\partial \hxyde}
         \Delta_\x \hxyde \\
         =  \frac{1}{3!}\mu^{A B C}_\x  \frac{\partial n^\x_{A B C}}{\partial \hxyde} 
         \left[ \frac{\partial X^D_\x}{\partial x^a} \frac{\partial X^E_\y}{\partial x^b} 
         \left[\delta g^{a b} - 2\nabla^{(a} \fd^{b)}_\x  \right] + g^{a b} 
         \frac{\partial X^D_\x}{\partial x^a}  \left(\lefx - \lefy\right) 
         \frac{\partial X^E_\y}{\partial x^b} \right] \\
         = \frac{1}{2} \dXYab \delta g^{ab} - \frac{1}{2} \dXYab \left(  \nabla^a \xi_\y^b +
         \nabla^b \xi_\x^a \right) + \rHXYa \left( \xi_\x^a - \xi_\y^a \right)   \ , 
\end{multline}
where
\beq
\dXYab = \frac{1}{3} \mu^{ABC}_\x 
          \frac{\partial n^\x_{ABC}}{\partial \hxyde}  \frac{\partial X^D_\x}{\partial x^a}  
            \frac{\partial X^E_\y}{\partial x^b} \ , 
\eeq
such that
\beq
\dXYab u_\x^a = 0 \ , \qquad \dXYab u_\y^b = 0  \ , 
\eeq
and
\beq
 \rHXYa = \frac{1}{3!}  \mu^{ABC}_\x \frac{\partial n^\x_{ABC}}{\partial \hxyde} \left(  
 g^{b c} \frac{\partial X^D_\x}{\partial x^b}  
 \nabla_a \frac{\partial X^E_\y}{\partial x^c} \right)\ .
\eeq
which is (notably) not guaranteed to be orthogonal to $u_\x^a$.

Finally, we have
\begin{multline}
 \frac{1}{3!}\mu^{A B C}_\x \frac{\partial n^\x_{A B C}}{\partial \hyde} \Delta_\x \hyde \\
    = \frac{1}{3!}\mu^{A B C}_\x \frac{\partial n^\x_{A B C}}{\partial \hyde} \left[ 
       \frac{\partial X^D_\y}{\partial x^a} \frac{\partial X^E_\y}{\partial x^b}
       \left(\delta g^{a b}- 2 \nabla^{(a} \fd^{b)}_\x \right)
       + g^{ab} \left(\lefx - \lefy\right) \left( \frac{\partial X_\y^D}{\partial x^a} 
       \frac{\partial X_\y^E}{\partial x^b}  \right) \right] \\
       = \frac{1}{2}\DXYab  \delta g^{a b} - \DXYab \nabla^{a} \fd^{b}_\y + \RHXYa 
       \left( \xi_\x^a - \xi_\y^a \right) \ ,  
\end{multline}
where we have used
\beq
 \DXYab =  \frac{1}{3} \mu^{A B C}_\x \frac{\partial n^\x_{A B C}}{\partial \hyde}   
                    \frac{\partial X^D_\y}{\partial x^a} 
                    \frac{\partial X^E_\y}{\partial x^b} = \DXYba \ , 
                    \eeq
and
\beq
  \RHXYa = \frac{1}{3!}  \mu^{A B C}_\x \frac{\partial n^\x_{A B C}}{\partial \hyde}   
                     \nabla_a \left(  g^{bc}\frac{\partial X^D_\y}{\partial x^b}
                     \frac{\partial X^E_\y}{\partial x^c}  \right) \ . \\
\eeq
In this case it is clear that neither $\DXYab$ nor $\RHXYa$ need to be flowline 
orthogonal with respect to $u_\x^a$ (although the former  is obviously  orthogonal to 
$u_\y^a$).

Putting all the results together, we arrive at 
\begin{multline}
      \mu^\x_a \delta n^a_\x = \mu^\x_a \left(  n^b_\x \nabla_b \fd^a_\x - 
              \fd^b_\x  \nabla_b n^a_\x -  n^a_\x \nabla_b \fd^b_\x \right) + \DXab\nabla^b  
              \xi_\x^a \\
             +  \sum_{\y \neq \x} \left[ \left( \RXYa + \rHXYa + \RHXYa \right)
              \left(\fd_\y^a - \fd_\x^a \right)  + \frac{1}{2} \dXYab \left( \nabla^a \xi_\y^b + 
              \nabla^b \xi_\x^a \right) + \DXYab  \nabla^{a} \fd^{b}_\y \right] \\
             + \frac{1}{2} \left[ \mu^\x_c n^c_\x g^{ab} + \DXuab + \sum_{\y\neq\x} 
             \left( \dXYusab + \DXYuab  \right) \right] \delta g_{ab}  \ . 
             \label{mudnxfinal} 
\end{multline}
Using
\beq
      \sum_{\x} \sum_{\y \neq \x} \dXYab \left(\nabla^a \xi_\y^b + \nabla^b \xi_\x^a \right) 
      = \sum_{\x} \sum_{\y \neq \x} \left(\dXYab + \dYXba\right) \nabla^b \xi_\x^a \ ,
\eeq
this means that Eq.~(\ref{varladiss}) becomes
\begin{multline}
    \delta \left(\sqrt{- g} \Lambda\right) = - \sqrt{- g} \left[\sum_{\x} \left(f^\x_a 
    + \Gamma_\x \mu^\x_a + \nabla^b \dtotxba - \rtotxa\right)  \fd^a_\x \right.  \\
   \left.- \frac{1}{2} \left(\Psi  g^{a b} + \sum_{\x} n^a_\x \mu^b_\x +  
    D^{a b}\right) \delta g_{a b}\right]  \ , \label{varlamdiss}
\end{multline}
where the dissipation tensor of the $\x$-component is 
\beq
    \dtotxba = \DXba + \sum_{\y \neq \x} \left[\DYXba + \frac{1}{2} \left(\dXYab + \dYXba
    \right)\right] \ , \label{disstotx}
\eeq
while the dissipative stress entering the stress-energy tensor is given by the sum
\beq
    D_{a b} = \sum_\x D^\x_{(a b)} = D_{b a} \label{totdisssym}
\eeq
(using the fact that the metric is symmetric). \ Finally, the total ``resistivity'' current is given 
by
\beq
    \rtotxa = \sum_{\y \neq \x} \left[ \left(\RYXa  - \RXYa\right) + \left( \RHYXa - 
                   \RHXYa \right) + \left( \rHYXa - \rHXYa\right) \right]\ .   
\eeq
It is important to note that these quantities still satisfy (as in the simpler model from 
Section~IIIC) 
\beq
      u_\x^a D^\x_{b a} = 0 \ , \label{dxabcons}
\eeq
and
\beq
  \sum_\x \rtotxa = 0 \ . \label{rvan}
\eeq

The final equations of motion  take the same form as before;
\beq
    f^\x_a + \Gamma_\x \mu^\x_a + \nabla^b \dtotxba =  \rtotxa 
             \ . \label{xfleqn2}
\eeq
The model is, however, richer. To demonstrate this, we return to the heat problem one 
final time.

\subsection{A Final Example}

Having introduced a number of additional dissipation channels, it is interesting to ask 
how the matter-heat problem changes. As before, we assume that $\Gamma_\n= 0$. 
From \eqref{rvan} it is also clear that we still have $R^\n_a=-R^\s_a$. 

As the matter component is conserved, we have 
\beq
{u_\n^a \nabla^b D^\n_{ba} = u_\n^a R^\n_a = - u_\n^a R^\s_a \quad \longrightarrow 
\quad D^\n_{ba}  \nabla^b u^a =  u^a R^\s_a \ ,} 
\label{mateq}
\eeq
where we have made use of \eqref{dxabcons}. This obviously reminds us of  \eqref{rel1}, 
but in this more general case $R^\n_a$ is not required to be orthogonal to either flow. If 
we use the decomposition
\beq
R^\n_a = R_u u_a + R_w w_a + \epsilon_{abcd} \phi_\n^b u^c w^d \ , 
\eeq
then \eqref{mateq} only involves $R_u$.  Moreover, $D^\n_{ba} $ is no longer required 
to be symmetric so there will now be a coupling to the antisymmetric piece 
$\varpi^\x_{a b}$ of the flow, see \eqref{deco}.

Turning to the entropy component, we have
\begin{multline}
T\Gamma_\s = - u_\s^a R^\s_a - D^\s_{ba} \nabla^b u_\s^a = \gamma \left( R_u - w^2 
R_w\right) - D^\s_{ba} \nabla^b u_\s^a \\
=  - \gamma w^2 R_w - \gamma D_{ba} \nabla^b u^a- \gamma D^\s_{ba} \nabla^b w^a   
\ , 
\end{multline}
where we have made use of the matter equation \eqref{mateq}. It is interesting to 
compare this final relation for the entropy creation rate to the corresponding results in 
the Newtonian case.  From Eq. (3.4) in \cite{Andersson10:_causal_heat} we have (in the 
notation of that paper)
\beq
T \Gamma_\s = - f^\n_i w_{\n\s}^i - D_{ij} \nabla^j v_\n^i - D_{ji}^\s \nabla_j w_{\n\s}^i \ .
\eeq
The salient features of the two relations are clearly the same. For example, the second 
law only constrains the resistivity along the relative flow. In addition, while the total 
dissipative stress tensor is symmetric (the second term in each relation) the individual 
contributions are not (the third term). This observation, which tends to be overlooked, is 
important as it links the entropy creation to the vorticity. 

At this point, it would be natural to develop a relativistic version of the Onsager argument 
used in \cite{andersson05:_flux_con,2011IJMPD..20.1215A,2012PhRvD..86f3002H}. 
This involves identifying  thermodynamic forces and fluxes, introducing an expansion 
with respect to an equilibrium state and making use of the relevant symmetries among 
the introduced coefficients.  We will postpone this step for a future effort. It is natural to 
do so because, so far we have not actually introduced the notion of an equilibrium state 
and we have certainly not based our analysis on an expansion away from such a state. 
In other words, the formalism we have developed is still general and 
nonlinear.\footnote{At first sight, this might seem somewhat peculiar and at odds with the 
basic principles of thermodynamics. However, it is useful to keep in mind that the 
(conservative) Euler equations are nonlinear (in the same sense as the present model) 
and do not have an immediate connection with the notion of an expansion away from an 
equilibrium. Such a connection may exist, although it would have to be at a deeper 
level.} It is perhaps a tribute to the elegance of the variational argument that we 
managed to get this far without taking what is often seen as one of the first steps of the 
analysis. However, the Onsager-type argument requires sacrifices and we will be forced 
to introduce a formal expansion in order to proceed. This will require some care. Further 
reason for caution comes from the fact that we are  working in spacetime. This means 
that the expansion of the different dissipative contributions will be more complex than in 
the three dimensional case. There are additional permissible forces/fluxes and the 
differential structure is richer. In contrast to the Newtonian case, one must  tread 
carefully as there are causality issues to consider. For all these reasons, it is natural to 
take a break at this point and return to the problem later. 

\section{Discussion/speculation}
\label{conclude}

We have presented an action principle for general relativistic multi-fluid systems 
including dissipation. The usefulness of the new formalism is that it can, at least in 
principle, circumvent ad hoc arguments often used in the traditional approach to the 
problem of dissipation in general relativistic fluid systems. Admittedly, this may not be
{\em the definitive} way to incorporate dissipation, but the new scheme is at least 
coherent and the line of reasoning is conceptually clear. The extension to more complex 
systems also seems relatively straightforward. This should be an advantage for 
astrophysical applications which are involving more detailed physics. For example, the 
coupling to electromagnetism is unambiguous, involving the usual minimal coupling 
Ansatz \cite{2012PhRvD..86d3002A}, and it should also be straightforward to account 
for issues involving polarisable media (although the details remain to be worked out). 
When it comes to neutron star models, it may be a matter of ``turning the crank'' to 
incorporate the elasticity of the outer crust
\cite{carter06:_crust,Samuelsson07:_axial_crust,2009CQGra..26o5016S}. Issues 
involving anisotropic lattices ought to be easy to accommodate, but extensions to 
models including say plastic flow still represent a challenge. 

It is, of course, not the case that the variational model is complete at this stage. 
Eventually we would like to turn the proposal into a plug-and-play scheme for relevant 
applications, but first we need to carry out a careful comparison between our model and 
the various alternatives. At the same time, one may speculate about potential 
extensions. In this final section we consider various issues that one might want to 
consider in more detail and suggest directions in which the model may be extended in 
the future.

\subsection{Completing the model}

The most pressing issue concerns the relationship between the variational model and 
its various predecessors. It is natural to ask to what extent the new formulation contains 
the same information regarding the possible dissipation channels as, for example, the 
celebrated Israel-Stewart construction 
\cite{Israel76:_thermo,Israel79:_kintheo1,Israel79:_kintheo2}. A comparison between 
the two descriptions may seem straightforward, but is in fact not trivial. That this is the 
case becomes apparent as soon as we note that the variational derivation did not 
involve an explicit expansion with respect to thermal equilibrium. In fact, we were never 
required  to consider possible equilibrium states at all. This is in sharp contrast to the 
usual approach that  takes an equilibrium  as its starting point and then carries out a 
formal expansion in terms of deviations from this state. There is, of course, nothing that 
prevents us from expanding the variational results in a similar fashion. In fact, in most 
situations of practical relevance this may be precisely what one ought to do.

It would seem natural to base such a construction on the standard Onsager approach 
\cite{onsager31:_symmetry}. This scheme was developed many decades ago, and 
provides a systematic formalism for determining the number and structure of dissipation 
channels of a given system. It also sheds light on how these can be woven together 
so that the second law is guaranteed to be satisfied. We have already considered this 
approachin detail in the Newtonian regime \cite{andersson05:_flux_con,2011IJMPD..20.1215A,2012PhRvD..86f3002H}, 
and would expect to draw on those results to guide us in the general relativistic context. 

Even though it derives from a powerful mathematical framework, we must remember that 
the variational model  is phenomenological. In order to apply it to physical systems, we 
need to connect the macroscopic model with a microscopic analysis. Such a model is 
required to provide the various transport coefficients, like the thermal conductivity and 
the various relaxation times. The standard approach to this problem is to resort to kinetic 
theory, building on a moment expansion for given velocity distributions together with an 
evaluation of the relevant collision integrals 
\cite{2011JPhCS.270a2042M,2013PhLB..720..347J}. More recent developments, which 
may be particularly relevant in the present context since the underlying Lagrangian for 
the theory is taken as starting point, derive the fluid dynamics from a field theory point of 
view \cite{2013PhLB..720..347J}. Future work needs to explore the connection between 
our new formulation and those efforts. 

It would seem natural to develop the link between the matter-space view of the present 
analysis and the coarse graining of phase space in statistical physics (see 
\cite{2011arXiv1110.6152V} for a potentially relevant discussion). It then becomes 
relevant to ask at what level the statistics should be considered. Is it at the spacetime 
level, or is it in the lower-dimensional configuration space? In principle, both answers 
seem viable but the latter would be an attractive (possibly quite revolutionary)
solution. In analogy with the description of elastic matter 
\cite{1972RSPSA.331...57C,carter73:_elast_pertur,2003CQGra..20..889B,1992GReGr..24..139M,2003CQGra..20.3613K,2012CQGra..29a5005G} 
one may envisage a model based on the notion of an evolving ``thermal geometry'' (in 
matter space) directly linked to the entropy change between hypersurfaces in 
spacetime. The model also requires dynamical map between matter space and 
spacetime, in order to link the changes in the local geometric structure of the matter 
configuration (described in terms of normal coordinates, say) to the macroscopic 
evolution of the system.

\subsection{Thermodynamical evolution}

Since the variational construction does not rely on an expansion away from 
thermodynamic equilibrium, it retains nonlinearities that may be relevant for a range of 
considerations. This may lay a foundation for a deeper understanding of nonlinear non-
equilibrium thermodynamics and in the extension lead to a framework to discuss the 
flow of time. As a starting point one might want to establish to what extent the variational 
model has an interior sense of time, e.g. associated with the constrained entropy 
evolution. If time is an emergent phenomenon, how does it depend on the imposed 
conditions? This is, obviously, a rather deep question but it is clear that the ``coordinate-
free'' representation of the variational approach provides an interesting starting point for 
a discussion of such foundational issues.

In the case of a dynamical evolution of a  general relativistic system, one must consider 
the role of different observers. This is a non-trivial issue in thermodynamics, closely 
related to the nature and interpretation of the entropy. Progress on this problem may 
require experience with the various formalisms for numerical relativity. In fact,  one might 
think that a variation of the 3+1 formalism would be natural in order to represent the 
internal clock of a system out of equilibrium. Building on the standard framework, one 
could consider to what extent the spacetime foliations are constrained by the 
thermodynamics. Are there a set of preferred observers imposed by (say) the entropy 
flow?

In addition to exploring issues concerning the foliation of spacetime, it would make 
sense to consider other physical constraints on the model. It is important to establish to 
what extent a matter/entropy model is constrained by fundamental principles. Take 
stability as an example. One would obviously expect any physical equilibrium model to 
be stable. Yet, at the same time one would want a system to exhibit instabilities in order 
to develop structures. This is another challenging problem. It is important to understand 
the difference between unphysical instabilities and ones that are expected in a realistic 
model. Building on the variational model, it would be interesting to investigate the 
various instabilities that this system exhibits. In connection with this, it would be natural 
to consider the role of the various energy conditions of general relativity. One would 
certainly want to understand how these conditions affect the thermodynamics and 
whether they constrain the evolution of a system.

\subsection{Hamiltonian formulation --- towards quantum aspects}

The consideration of boundary terms (ignored throughout our discussion) in the 
variational approach suggests that an alternative approach to the problem may prove 
useful.  The boundary forms a spacelike two-surface, which will have associated to it
two null directions orthogonal to the surface. By considering the extrinsic curvature of 
the two-surface in these directions  one can invariantly define ingoing and outgoing null 
vectors which implies that the physics may be naturally represented by a 2+2 foliation of 
spacetime. This formulation has a clear geometric interpretation and it would make 
sense to explore analogous ideas for the description of thermodynamics. In doing so, 
we expect to compare and contrast different approaches to the spacetime evolution 
problem in order to establish the most natural framework for thermodynamical 
evolutions. 

This research direction is entirely within the realm of classical physics. Yet, one would 
ultimately need to account for quantum aspects. While we make progress at the 
classical level, we should prepare the ground for future explorations of the quantum 
arena. In absence of a theory for quantum gravity this is obviously a huge step, but some 
basic principles seem clear. For example, if we want to discuss quantization then we 
need to develop a Hamiltonian description for the relativistic thermo/hydrodynamics. 
This is known to be a challenging problem, but Dirac's procedure for developing a 
Hamiltonian system from a given Lagrangian is (at least in principle) clearly laid out.
However, the steps involved are far from straightforward in practice. This is particularly 
true in the case of a constrained variational model, as in the present case. Nevertheless, 
one should be able to map out this important problem by considering in detail the 
involved first- and second class constraints.

\subsection{Fundamental physics}

That  gravity and thermodynamics are intimately linked is clear from from the 
equivalence of energy and mass, which implies that heat must affect the gravitational 
field. Nevertheless, from a conceptual point of view it is not understood to what extent 
the gravitational field is ``hot''. Basically, we do not yet have an operational definition of 
the entropy associated with an evolving gravitational field \cite{2011PhRvD..83d4048P}. 
This is a long-standing problem. The variational model for heat accounts for the 
coupling to (and evolution of) the gravitational field via the Einstein field equations. 
However, so far the main focus has been on the matter sector of the problem. It would be 
interesting to broaden the discussion and explore the role of the gravitational field in 
more detail. The aim would be to establish how the second law of thermodynamics 
feeds into the evolution of the gravitational field, and (conversely) how variations in the 
gravitational field affect the entropy and the heat flow. A conceptually interesting  issue 
concerns the link between observers and the increase of gravitational entropy, and the 
(obvious) link to the microstates associated with a black hole's event horizon. 

The variational approach  provides the foundation for the exploration of a range of  
relevant issues. Let us  comment on three, perhaps particularly topical, issues;

\begin{enumerate}
\item
 The 2+2 approach (discussed above) has interesting connections
with the Ashtekar formulation of quantum gravity, using the description due
to Jacobson and Smolin based on self-dual 2-forms \cite{jac1,vic1,vic2}.  The key
feature of using a null foliation is that the Hamiltonian constraint is no
longer first class and the remaining first class contraints form a Lie
algebra \cite{vic3}. 

\item
The notion of a two-dimensional boundary is obviously very similar to the ideas behind 
gauge/gravity duality and the holographic principle. In fact, the use of a null foliation 
plays a key role in Jacobson's derivation of the Einstein equations from thermodynamic 
principles \cite{jac2}. Given this, it would make sense to develop the connection with 
dissipative holographic fluid dynamics. Many such models consider the fluid limit of 
conformal field theories, starting from a suitable Lagrangian and generating dissipative 
terms by a formal derivative expansion \cite{2009CQGra..26v4003R,2011JHEP...04..125B,baier}. 
This has led to the identification and exploration of dissipative terms that were not 
present in the classic Israel-Stewart construction. This is exciting progress, but it is 
important to keep in mind that much of the holographic-fluid program has so far  
focussed on rather unphysical models, e.g. systems with conformal symmetry. 
Nevertheless, the discussion  promises  a deeper understanding in the future, and it 
would make sense to  investigate the connection between our new dissipative 
variational approach and the quantum field theory-led holography models in detail. This 
is particularly interesting since the gauge/gravity approach may provide insight into the 
microphysics origin of the various dissipation channels.   

\item
The Hamiltonian formulation of the problem 
\cite{1971PhRvD...4.3559S,1979agn..book..273C,1984PhLA..101...23H,1985PhyD...17....1H,comer93:_hamil_multi_con,comer94:_hamil_sf} 
would allow us to make direct contact (and build upon) the notion of thermal time 
\cite{connes,2013PhRvD..87h4055R} (associated with the evolution of pre-symplectic 
systems). So far, this concept has been developed for systems in thermal equilibrium 
(for which there is a clear description). It is relevant to ask how the concept is altered by 
non-equilibrium effects. This is a natural problem to consider given that the 
thermodynamic arrow of time relies on the second law (irreversibility), and hence 
``applies'' only to non-equilibrium systems. 
\end{enumerate}

This list of topics and issues formulates an ambitious research programme based on the 
new variational model for dissipative systems. Some of the problems are clearly 
achievable, and one might expect to make swift progress on them. Other problems are 
more speculative and foundational in nature. These targets may be much harder to 
reach, but at least the new model provides a fresh approach that may lead to the 
development of interesting perspectives.

\acknowledgments
We are grateful to Ian Hawke and James Vickers for useful discussions and for 
commenting on early versions of this work. NA acknowledges support from STFC 
through grant number ST/J00135X/1.  GLC acknowledges partial support from 
NSF via grant number PHYS-0855558.

\end{document}